\documentclass[12pt]{article}

\usepackage{amsthm,amsmath} %
\usepackage[square,sort,comma]{natbib}

\RequirePackage[colorlinks,citecolor=blue,urlcolor=blue]{hyperref}

\usepackage{cleveref}
\usepackage{stackrel}
\usepackage{booktabs}
\usepackage{makecell}
\usepackage{amsthm}
\usepackage{amsmath}
\usepackage{amssymb}
\usepackage{bbm}
\usepackage{mathtools}
\usepackage{floatrow}
\usepackage{authblk} 
\usepackage{caption} 
\usepackage{geometry} 
\usepackage[table,x11names]{xcolor} 
\usepackage{hyperref}

\usepackage{chngcntr} 
\counterwithin*{section}{part}

\usepackage[utf8]{inputenc}
\usepackage{chngpage} 
\usepackage{subcaption} 
\usepackage[pdftex]{graphicx}

\usepackage{float}
\floatstyle{plaintop}
\restylefloat{table}

\newcommand{\indep}{\rotatebox[origin=c]{90}{$\models$}}
\crefname{section}{§}{§§}
\Crefname{section}{§}{§§}

\numberwithin{equation}{section}
\theoremstyle{plain}

\newtheorem{prop}{Proposition}  
\newtheorem{lemma}{Lemma}  

\theoremstyle{definition}
\newtheorem{defn}{Definition}[section]

\providecommand{\keywords}[1]{\textbf{\textit{Keywords:}} #1}
\renewenvironment{abstract}
{\small
	\begin{center}
		\bfseries \abstractname\vspace{-.5em}\vspace{0pt}
	\end{center}
	\list{}{%
		\setlength{\leftmargin}{0mm}
		\setlength{\rightmargin}{\leftmargin}%
	}%
	\item\relax}
{\endlist}

\newcommand{\bX}{{\bf X}}
\newcommand{\bY}{{\bf Y}}
\newcommand{\bb}{{\bf b}}
\newcommand{\be}{{\bf e}}
\newcommand{\bP}{{\bf P}}
\newcommand{\bj}{{\bf j}}

\title{Testing for Differential Abundance in Compositional counts data, with Application to Microbiome Studies}

\author[*]{Barak Brill}
\author[**]{Amnon Amir}
\author[***]{Ruth Heller}
\affil[*]{Tel Aviv University, Tel Aviv, email for correspondence: \href{mailto:barakbri@mail.tau.ac.il}{barakbri@mail.tau.ac.il}}
\affil[**]{Sheba Medical Center, Tel Hashomer, affiliated with the Tel Aviv University}
\affil[***]{Tel Aviv University, Tel Aviv, email for correspondence: \href{ruheller@gmail.com}{ruheller@gmail.com}}

\begin{document}
\newgeometry{top=20mm, bottom=20mm} 
\bibliographystyle{plainnat}
\maketitle

\begin{abstract}
	Identifying which taxa in our microbiota  are associated with traits of interest is important for advancing science and health. However, the identification is challenging because the measured vector of taxa counts (by amplicon sequencing) is compositional,  so a change in the abundance of one taxon in the microbiota induces a change in the number of sequenced counts across all taxa. The data is typically sparse, with zero counts present either due to biological variance or limited sequencing depth (technical zeros). For low abundance taxa, the chance for technical zeros is non-negligible. We show that existing methods designed to  identify differential abundance for compositional data may have an inflated number of false positives due to improper handling of the zero counts.  We introduce a novel non-parametric approach which provides valid inference even when the fraction of zero counts is substantial. Our approach uses a set of reference taxa that are non-differentially abundant, which can be estimated from the data or from  outside information.  We show the usefulness of our approach via simulations, as well as on three different data sets: a Crohn's disease study, the Human Microbiome Project, and an experiment with 'spiked-in' bacteria.

A \verb|R| software package, \verb|dacomp|, implementing the novel methods suggested is publicly available.
\end{abstract}

\restoregeometry
\section{Introduction}\label{sec:Introduction}
The microbiome is the collection of micro-organisms and bacteria which are part of the physiological activity of a host body or ecosystem \citep{hamady2009microbial}. It is of 
interest to associate change in microbial structure to disease and other environmental factors.  
For example, the study of \citet{vandeputte2017quantitative} investigated the change in the microbial ecology of fecal samples, in the presence of Crohn's disease. This change is associated with a change in the composition of the gut microbiome of patients. A better understanding of the microbial changes in the gut may lead to a better understanding and treatment of Crohn's disease. 

A common method of measuring the composition of the bacterial community is by sequencing the 16S rRNA gene, which codes for a crucial part of the ribosome common to all living cells. The variable regions in the 16S rRNA gene 
are subject to mutations along genetic lineages. Due to these variations, 16S rRNA sequence patterns serve as a proxy for the taxonomic identification of their organism.


The data is generated by collecting samples from different specimen, and the targeted variable regions are duplicated and amplified using PCR. Sequencing technology allows one to read the amplicons of the PCR procedure and list all sequences read for each sample. This list of sequences is then trimmed to a constant length of, e.g., 150 base pairs \citep{nelson2014analysis}, and the amount of each unique sequence in each sample is recorded. Due to errors in the sequencing, not all unique sequences actually represent unique bacteria. In order to identify the bacteria actually present in each sample, two alternative methods can be used: Operational taxonomic units (OTUs) are sequences which differ up to a certain threshold, e.g., 3\% of base pairs out of 150 \citep{hamady2009microbial}. Amplicon sequence variants (ASVs) are the individual  sequences \citep{callahan2016dada2,amir2017deblur} obtained after a denoising of the reads. The OTUs, or ASVs,  represent the finest resolution of organism type identifiable from sequencing variants of the 16S rRNA gene. The data therefore consists of the  number of observed sequences for each OTU or ASV in each sample. 
The units of interest  for analysis, referred to as taxa, are the single OTUs or ASVs, or coarser units that aggregates phylogenetically related OTUs or ASVs, e.g., into genera.  


Several challenges are encountered when trying to identify which taxa are associated with a trait based on the observed counts per taxon. The first challenge is that the number of sequenced reads, or \textit{sequencing depth}, varies from sample to sample, and is mostly an artifact of the sequencing procedure rather than a proxy to the sample's original abundance of bacteria, also known as the sample's \textit{microbial load}. Therefore, only the  relative frequencies are informative, i.e., the count data is compositional 
\citep{gloor2017microbiome,mandal2015analysis,kumar2018analysis}.

The second challenge is that the vector of taxa counts is sparse by nature, as not all taxa are measured in all samples. The percentage of zeros in the data ranges between 50\% and 90\%  for many types of samples \citep{xu2015assessment}. A taxon with zero counts can occur for two reasons: (1) low frequency in the sampled units, so the sample does not capture the very rare taxa, henceforth referred to as technical zeros; (2) taxa not shared by the entire population, henceforth referred to as structural zeros.

Additional challenges are the study size (the number of samples can be much smaller than the number of taxa, \citealt{nelson2014analysis}), and the strong (yet unknown) dependence between taxa counts. Intuitively, compositionality implies negative correlations. However, strong positive correlations between taxa across subjects are also observed \citep{hawinkel2017broken}. 

Due to the above challenges, it is difficult to design a valid inferential method for identifying the taxa that are associated with the trait. 
Statistical tests that ignore compositionality can lead to false positive findings
, as demonstrated by the following example.

\emph{Example 1: a toy example demonstrating the danger of ignoring compositionality.}  Suppose we have a binary trait, i.e., two groups of samples. The vector of counts for each sample is multinomial with $N$ total counts,  and the probability vector is, for a sample that belongs to the first group, $\bP$, and for a sample that belongs to the second group,   $(1-w)\times \bP + w\times\be_1$, where $\be_1$ is the binary vector with a single entry of one in the first coordinate   
and $w\in\left(0,1\right)$. Since 
the first taxon has an increased relative frequency in the first group compared to the second group, and all other taxa have decreased relative frequency in the second group compared to the first group, 
for large enough sample sizes, the two-sample test for equality of relative frequencies will reject the null hypothesis at each coordinate. However, we are interested in detecting only the first taxon, since it is the only one driving the observed differences across groups. (In microbiome studies, unlike in this example, the probability vector varies within each group.)

In this paper, our goal is to develop a method for statistical inference in a compositional setting which considers as true discoveries only the taxa whose original ecosystem abundance has changed. The original ecosystem abundance of taxa cannot be reconstructed from their relative frequencies. However,  a change in the absolute abundance of a taxon may be detectable 
with respect to a reference frame of taxa \citep{morton2019establishing}.

In  \S~\ref{subsec:existing_methods}, we review methods for analysis of differential abundance in microbiome studies and point out limitations which this work aims to overcome. 
In  \S~\ref{sec:Notations}, we formalize our analysis goal of detecting differential abundance. In \S~\ref{sec:valid_approach} we describe our main result, a testing procedure for discovering the differentially abundant taxa that has guaranteed control over false positives. 
This test relies on the availability of a  reference set of taxa, and we  show how to estimate this reference set from the data. In \S~\ref{sec:simulations} and \S~\ref{sec:RDE_GUT} we compare the performance of our method against other methods in, respectively, simulations and real data examples. 
In  \S~\ref{sec:Discussion} we conclude with final remarks. 

\subsection{Review of methods for differential abundance analysis}\label{subsec:existing_methods}
Let $\bf{X}$ be the $m$-dimensional vector of observed taxa counts. Let $C(\bX): \mathbb{R}^m \rightarrow \mathbb{R}^+$ be a normalization function, so that the analysis will associate the scaled counts, $\bX / C(\bX)$, with the trait. 
Total sum scaling (TSS) normalization selects $C(\bX)$ to be the total number of counts in $\bf X$. Example~1  demonstrates that the trait may be associated with  a non-differentially abundant normalized taxon, so a test of independence following normalization cannot be used to identify the differentially abundant taxa.  

\citet{paulson2013differential} suggested cumulative sum scaling (CSS). CSS normalization selects $C\left(\bX\right)$ so that the smallest $q_{CSS}$ values in $\bX$ sum to one, with $q_{CSS}$ chosen adaptively from the data. As with TSS, this normalization does not resolve the bias in testing induced by compositionality, as shown in  \cite{mandal2015analysis}, as well as in our simulations in  \S~\ref{sec:simulations}. 

Other scaling and transformation methods can be found in 
\citet{kaul2017zeros}, which adapted the normalization of \citet{aitchison1986statistical} for use in microbiome studies. But after transformation, the null hypothesis of independence between  a  taxon and the trait will be false also for non-differentially abundant taxa since the scaling factors considered are  functions of the differentially abundant taxa. 
Even taking $C(\bX)$ to be the number of counts in a specific taxon, e.g. the $m$th taxon, is problematic since typically for every taxon some samples will have a zero count, so a pseudo-count has to be put in place of zero. If the probability of zero counts changes with the trait, then 
$C(\bX)$ is associated with the trait. 

\citet{kumar2018analysis} suggested an alternative scaling approach, called Wrench, based on the assumption that taxa not associated with the condition of interest have maintained the ratios of their respective proportions in each sample. To briefly describe the approach, we make use of the setup presented in Example 1. \citet{kumar2018analysis} observe that while the expected values of all coordinates differ across study groups, coordinate means across all taxa except the first taxon were lowered in the second group compared to the first group by the same multiplicative factor. In Example 1, this ratio is given by the multiplier $1-w$. \citet{kumar2018analysis} suggest estimating the common multiplicative factor from the data for scaling taxa counts.


\cite{fernandes2013anova} suggested the  ALDEx2 method and software package 
, where the normalization factor  $C(\bf{X})$, is taken to be the geometric mean of the counts observed in  a subset of the taxa. 
The counts are normalized  with respect to taxa that are estimated to be non-differentially abundant, and then log-transformed for statistical inference, as detailed in \cite{fernandes2013anova}. However, in order to avoid division by zero, a pseudocount of 0.5 is added to all data entries. 
If the probability of zero counts changes with the trait, then the inference may not be valid, but the bias is less severe than with the previous methods, as we show in \S~\ref{subsec:sim_gut}. 


Additional methods making use of auxiliary measurements to determine normalization factors include the approach of \citet{vandeputte2017quantitative}, which suggested the use of flow-cytometric measurements as a means to estimate the absolute microbial load of samples; the approach of \citet{stammler2016adjusting} which suggested artificially inserting bacteria of types non-endemic to the measured samples in predetermined abundance; and the use of spiked-in DNA sequences \citep{quinn2018field}.


\citet{mandal2015analysis} suggested a framework for analysis under composionality (ANCOM) which avoids the need of a "per-sample" scaling factor. The key, very reasonable,  assumption is that the effect of compositionaly is such that inter-taxa ratios are maintained for non differentially abundant taxa. For the two-sample case, the ANCOM procedure is as follows. Let $p_{j,k}$ denote the $p$-value obtained for the Wilcoxon rank sum test comparing the ratio between the the $j$th and $k$th taxa, across the two groups. ANCOM computes $p_{j,k}$ for every pair of taxa, $j,k$. In order to avoid division by zero, ANCOM adds a pseudocount with a value of $1$ to all counts values. Let the indicator function for $p_{j,k}$ being below or equal to a value of $\alpha$ be $I_{j,k} = \mathbbm{1}\left(p_{j,k} \leq \alpha \right)$. The number of pairwise rejections consisting of taxon $j$ is denoted by $\mathcal{W}_{j} = \sum_{k=1,k\ne j}^{m} I_{j,k}$. By assumption, frequencies of non differentially abundant taxa maintain their respective ratios, so in a well powered study it is expected that the number of rejections per taxon, $\mathcal{W}_{j}$, will be relatively high for the differentially abundant taxa. The taxa with indices $\{j| \mathcal{W}_{j} \geq \mathcal{W}^*\}$ are declared to be  differentially abundant, where $\mathcal{W}^*$ is chosen adaptively as detailed in  \citet{mandal2015analysis}.


Two related problems in all above normalization methods are (1) non-differentially abundant taxa remain associated with the trait if the prevalence of zero counts varies with the trait, since zero counts cannot be scaled; and (2) many of the methods, in order to apply transformations, use pseudo-counts instead of the zero counts, which corresponds to microbial load not measured in practice. We will demonstrate 
that the mishandling of zero counts in the above approaches can lead to an unacceptably inflated rate of false positive discoveries. 


\section{The setup and goal}\label{sec:Notations} 
We assume a general setup for the generation of taxa counts. 
Let $m$ and $n$ be the number of taxa and   samples, respectively. For sample $i\in \{1,\ldots, n\}$, we denote by $N_i$ the total number of counts sampled,  by $\bY_i$ the measured (univariate or multivariate) trait, and by $\bX_i$ the $m$-dimensional vector of observed taxa counts.  
Let $\bP_i$ be the (unobserved) vector of the taxa population relative frequencies  
in sample $i$. 
We assume  that $\bX_i$ is a  multinomial sample with parameters $N_i$ and $\bP_i$.  

For simplicity, we omit the sample subscript $i$ when addressing a single observation. 
For an $m$-dimensional binary vector with at least one entry of one, ${\bf s} = (s_1,\ldots, s_m)'$, we denote by $\bX({\bf s})$ and $\bP({\bf s})$ the subvectors of $\bX$ and $\bP$ of dimension ${\bf s}'{\bf s}$ (the number of nonzero entries in ${\bf s}$) containing the coordinates for which $s_i=1$. The sum of the entries in these subvectors is ${\bf s}'\bX$ and ${\bf s}'\bP$.
We denote by $\be_j$ the $m$-dimensional binary vector with a single entry of one at coordinate $j$, so $\bP(\be_j)$ and $\bX(\be_j)$ are the population relative frequency 
and observed count for taxon $j\in \{1,\ldots,n\}$. Our setup is thus that we observe $n$ realizations of $(\bX, \bY)$, where $\bX$ is a vector of multinomial counts given the (unobserved) random vector $\bP$ of population relative frequencies for the subject,   
$$\bX | \bP, N \sim multinom\left( N , \bP \right) ,\quad 
\bP(\be_j)\geq 0 \textrm{ for } j=1,\ldots, m, \quad {\bf 1}'\bP = 1.$$

We aim to identify the taxa that are associated with $\bY$, taking the compositional nature of the data into account. For this purpose, we assume that there exists a group of taxa that may be associated with $\bY$ via their sum, but are otherwise independent of $\bY$. Specifically, denoting the actual  abundance of the $m$ taxa for the observation  by ${\boldsymbol \mu}$, we have the relationship ${\boldsymbol \mu}/{\bf 1' \boldsymbol \mu} = \bP$. We assume that there is a subset vector ${\boldsymbol \mu}({\bf s})$ that is independent of $\bY$ except possibly through a change in the total sum ${\bf s}'{\boldsymbol \mu}({\bf s})$. The dependence on the sum may occur,  for example, if  an increase in other taxa (with relation to $\bY$) caused this subset of taxa to be less prevalent, but the relationship between the coordinates of this subset is unchanged with $\bY$. Therefore,  ${\boldsymbol \mu({\bf s})}/{\bf s' \boldsymbol \mu} = \bP({\bf s})/{\bf s}'\bP$ is independent of  $\bY$ (see  \citealt{mandal2015analysis} for a  similar assumption) . Such a group of taxa can serve as a {\it reference set}, defined below, for pointing towards the discoveries of interest. We use the symbol $\indep$ to mean that two random vectors  are mutually independent. 


\begin{defn}\label{defn-refset}
	A set of taxa with indices $ \{b_1,...,b_r\}$ is  a \emph{reference set} if for the $m$-dimensional indicator vector $\bf b$  with exactly $r$ ones  at  entries $\left(b_1,...,b_r\right)$,   ${\bf b}'\bP >0$ with probability one, and 
	\begin{equation}\label{eq:NullAssumption}
	\frac{\bP({\bb})}{{\bb}'\bP }\indep \bY.
	\end{equation} 
\end{defn}


Our goal is to find all taxa which are differentially abundant, i.e., 
taxa which vary with  $\bY$  given the reference set, while taking compositionality into account.  For  a given  \emph{reference set} of $r$ taxa, 
let $\bb_j$ be the $m$-dimensional binary vector with entries of one in $\{b_1,b_2,...,b_r\}$ and in $j$, where   $j\notin \{b_1,b_2,...,b_r\}$. 
The null hypothesis to be tested for a single taxon $j$
is that taxon $j$ is not differentially abundant:

\begin{equation} \label{eq:eqdist_latent}
H_0^{(j)}:\quad \frac{\bP(\bb_j)}{\bb_j'\bP}\indep \bY.
\end{equation}

If $H_0^{(j)}$ is false, then the normalized vector of relative frequencies which includes taxon $j$ and the reference set  varies with $\bY$ and this variability is not a consequence of a change in the relative abundance of the sum of the taxon and the reference set. Thus, we would like to identify all  taxa for which the null hypothesis in \eqref{eq:eqdist_latent} is false. 

More generally, we can consider testing a group of taxa together. Let $\bb_\bj$ be the $m$-dimensional binary vector with entries of one in $\{b_1,b_2,...,b_r\}$ and in $\bj$, where $\bj$ is a vector of indices satisfying  $\bj\cap \{b_1,b_2,...,b_r\} = \emptyset.$ The null hypothesis to be tested for a group of  taxa $\bj$
is that none of them are differentially abundant:

\begin{equation} \label{eq:eqdist_latent2}
H_0^{(\bj)}:\quad \frac{\bP(\bb_\bj)}{\bb_\bj'\bP}\indep \bY.
\end{equation}
If $H_0^{(\bj)}$ is false, then  the normalized vector of relative frequencies which includes $\bj$ and the reference set  varies with $\bY$ and this variability is not a consequence of a change in the relative abundance of the sum of the taxa in $\bj$ and the reference set.

We are unable to test these null hypotheses directly, since $\bP$
is not observed. Before proceeding to present our valid tests  in \S~\ref{sec:valid_approach}, we discuss  testing approaches that may appear natural, but are in fact non-valid when the data is overdispersed and has a nonnegligible amount of zero counts.

A simplified analysis may ignore  the fact that $\bP$ varies across observations, i.e., that the data is over-dispersed. This simplification allows application of well-known tests, but can severely affect the level of the test. Specifically, for a binary $\bY$, ignoring over-dispersion reduces to a test of whether $\bP(\bb_\bj)/\bb_\bj'\bP$ is identical across the two groups. In  \S~\ref{sec:simulations} we show that the level of the Fisher exact test in this case can be much higher than the nominal level. 

Another simplified analysis may replace the unobserved $\bP$ with the observed $\bX$ in the test of  \eqref{eq:eqdist_latent}, thus  rejecting \eqref{eq:eqdist_latent} if the test of  $\bX(\bb_j)/\bb_j'\bX \indep \bY $ is rejected. However, the distribution of  $\bX(\bb_j)/\bb_j'\bX$ depends on $\bb_j'\bP$, and $\bb_j'\bP$ may depend on $\bY$ even if \eqref{eq:eqdist_latent} is true. Therefore, even if  $\bX(\bb_j)/\bb_j'\bX $ and $ \bY $ are dependent, \eqref{eq:eqdist_latent} may be true. 

In analysis of compositional data, it is popular to add a pseudo count, since testing that $\bX(\bb_j)/\bb_j'\bX \indep \bY $ is possible only if $\bb_j'\bX$ is non-zero for all samples. We conclude this section with a numerical example that shows that the inflation in the level of the test $H_0^{\left(j\right)}$ with and without the additional of a pseudo-count can be non-negligible. The  inflation is larger with the addition of a pseudo count for each configuration examined, and  the inflation increases with larger differential abundance. 

\begin{figure}[htbp]	
	\centering
	\includegraphics[width=0.9\textwidth]{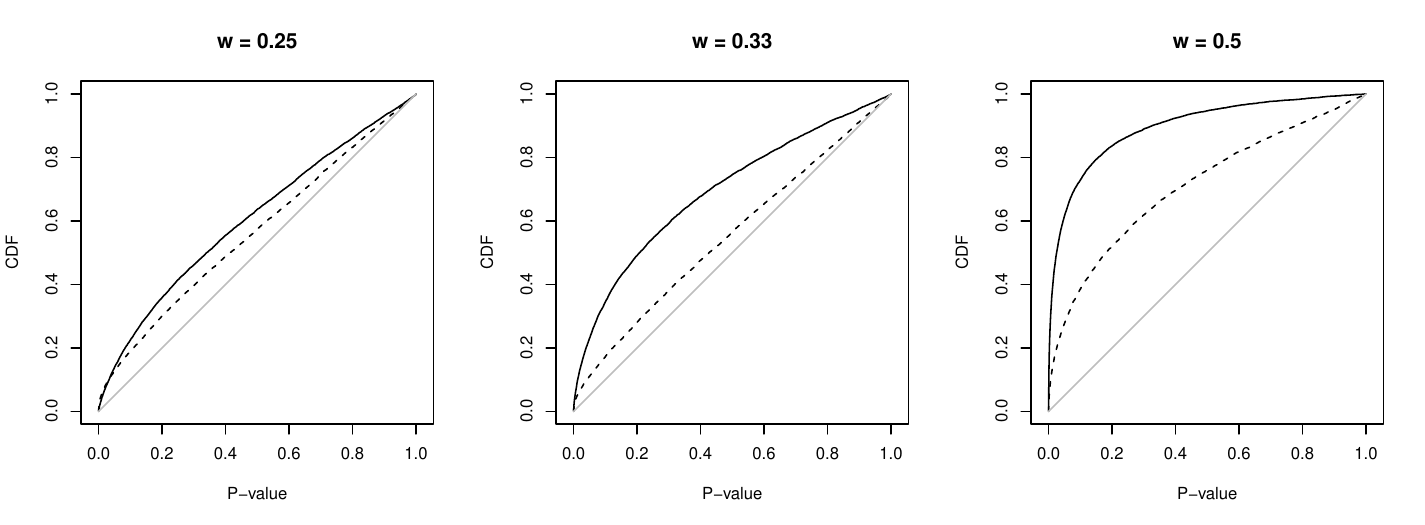}
	\caption{
		Cumulative distribution function (CDF) of the  $p$-value, for testing the independence between $\bX(\be_2)/\max\left(1,\bX(\be_2)+\bX(\be_3) \right)$ and $\bY$, in the setting defined by \eqref{eq:counter_example_relative_freq}, without a pseudocount (dashed) and with a pseudocount of 1 (solid), for $w\in \{0.25,0.33,0.5\}$. The CDF of the uniform distribution is in gray.
	}
	\label{figure:incorrect_approaches}
\end{figure}

\emph{ Example 2: the effect of using pseudocounts.} We consider a setting with $n=100$ samples and a constant sequencing depth of $N_i = 5000$ for all samples. The trait $\bY$ is binary, and the population relative frequencies of taxa are
\begin{equation}\label{eq:counter_example_relative_freq}
\bP = \left\lbrace
\begin{array}{ll}
\left(1-\frac{6}{N},\frac{1}{N},\frac{5}{N}\right)' & \text{if } \bY=0,\\
\left(1-w\right)\cdot \left(1-\frac{6}{N},\frac{1}{N},\frac{5}{N}\right)'+w\cdot \left(1,0,0\right)' & \text{if } \bY=1.
\end{array} \right.
\end{equation}
where $w\in (0,1)$. The parameter $w$ represents an increase in the total microbial load. For example, $w=0.25$ represent a $33\%$ increase in the total microbial load of samples in the group where $\bY=1$ compared to samples from group where $\bY=0$, resulting from an increase in the absolute abundance of taxon 1 alone. We test taxon 2 for differential abundance, with taxon 3 given as a reference. The null hypothesis $H_0^{\left(2\right)}$ is true. We use the Wilcoxon rank-sum test to test for equality of distribution  of $\bX(\be_2)/\max\left(1,\bX(\be_2)+\bX(\be_3) \right)$ across the two $\bY$ groups, with and without the addition of a pseudo count to  $\bX(\be_2)$ and $\bX(\be_3)$.
Table~\ref{table:incorrect_approaches_T1E} shows the unacceptably high type I error probability for  $w\in \{0.25,0.33,0.5\}$. Figure~\ref{figure:incorrect_approaches} shows that the distribution of the $p$-value is stochastically smaller than the uniform distribution, so it is not a valid $p$-value for $H_0^{\left(2\right)}$.

\begin{table}[htbp]
	\caption{ Probabilities for type I error when testing for independence between $\bX(\be_2)/\max\left(1,\bX(\be_2)+\bX(\be_3) \right)$ and $\bY$ using the Wilcoxon rank sum test at $\alpha=0.1$, in the setting defined by \eqref{eq:counter_example_relative_freq}, for 3 values of $w$. Based on $10^4$ simulations.
	}
	\label{table:incorrect_approaches_T1E}
	\centering
	\begin{tabular}{c|c|c|c}
		\hline 
		& $w=0.25$ & $w=0.33$ & $w=0.5$ \\ 
		\hline 
		no pseudocount & 0.19 & 0.17 & 0.38 \\ 
		pseudocount of 1 & 0.22 & 0.34 & 0.73 \\ 
		$P\left(\bX(\be_2)+\bX(\be_3) = 0|\bY=0\right)$ & 0.002 & 0.002& 0.002\\
		$P\left(\bX(\be_2)+\bX(\be_3) = 0|\bY=1\right)$ & 0.011 & 0.018 & 0.05\\
		\hline 
	\end{tabular} 
\end{table}

\section{Testing for differential abundance}
\label{sec:valid_approach}

\subsection{A valid test given a reference set of taxa} \label{subsec:oracle_test}
Let  $\{b_1,...,b_r\}$ be a reference set of taxa, as defined in \ref{defn-refset}, for sample $\{(\bX_i, \bY_i): i=1,\ldots, n \}$. In this section, we assume that the reference set is known and that  the total count in the reference set is positive for each observation, i.e., $\min_{i=1,\ldots,n} \bb'X_i(\bb)>0$. In \S~\ref{subsec:choosing_reference} we address the problem of estimating such a reference set. 

Our testing approach relies on a key observation that if the null hypothesis \eqref{eq:eqdist_latent} (or \eqref{eq:eqdist_latent2}) is true, then counts that are properly rarefied will be independent of the trait $\bY$. Therefore, rejection of the hypothesis of independence between these rarefied counts and $\bY$ will lead to rejection of \eqref{eq:eqdist_latent} (or \eqref{eq:eqdist_latent2}) with the desired nominal type I error level control guarantee.  

For simplicity, we describe our approach for testing the null hypothesis \eqref{eq:eqdist_latent}, but similar steps follow for the null hypothesis \eqref{eq:eqdist_latent2}. For taxon $j\in \{1,\ldots, m\}$, 
the conditional distribution of $\bX(\be_j)$ given $\bb_j'\bX$ and $\bP$ is binomial with parameters $\bb_j'\bX$ and $\bP(\be_j)/\bb_j'\bP$. Therefore, even if  \eqref{eq:eqdist_latent} is true, i.e., $\bP(\bb_j)/\bb_j'\bP\indep \bY$, $\bX(\be_j)$ depends on $\bY$ if $\bb_j'\bX$ depends on $\bY$. However, if we rarefy the counts for taxon $j$ to a depth $\lambda_j$,  by sampling from the hypergeometric distribution with parameters $\lambda_j, \bX(\be_j),$ and  $\bb_j'\bX$, then the distribution of the rarefied count is binomial with parameters $\lambda_j$ and $\bP(\be_j)/\bb_j'\bP$, see \S~\ref{app_A_proof_of_lemma} for details. Therefore, if  \eqref{eq:eqdist_latent} is true,  the rarefied count  for $\bX(\be_j)$ is independent of  $\bY$ even if $\bb_j'\bX$ depends on $\bY$. We thus suggest testing  \eqref{eq:eqdist_latent}  using the following procedure: 
\begin{enumerate}
	\item Compute the minimum total counts of the taxon and the reference set, $\lambda_j = \min_{i=1,\ldots,n} \bb_j'\bX_i. $
	\item For each observation $i=1,\ldots, n$, sample a count from the hypergeomtric distribution with parameters $\lambda_j, \bX_i(\be_j), \bb_j'\bX_i.$ The sampled count is denoted by ${\b Z}_{i, \lambda_j}(\be_j)$.
	\item Test the null hypothesis of independence between the rarefied count ${ Z}_{\lambda_j}(\be_j)$ and $\bY$ using an appropriate $\alpha$ level test for the data $\{({Z}_{i,\lambda_j}(\be_j), \bY_i): i=1,\ldots, n \}$. 
\end{enumerate}

\begin{prop}\label{prop-testvalidity}
	If the null hypothesis \eqref{eq:eqdist_latent} is true, then the aforementioned testing procedure has level $\alpha$. 
\end{prop}
See \S~\ref{app_A_proof_of_lemma} for a  proof. We note that our testing procedure assumes that  $\lambda_j>0$. In \S~\ref{sm-lambdapos-example} we provide an example that shows that removing the samples with zero (or a low) count results in a biased test. Therefore, for applying our testing procedure to  all $j\in \{1,\ldots,m\}/\{b_1,\ldots,b_r\}$, we require that  the reference set will have enough counts for each sample, so that $\min_{i=1,\ldots,n}\bb'\bX_i>0$ and hence $\lambda_j>0$. Samples with extremely low sampling depth (technical faults) may be removed from the entire analysis  if it is reasonable to assume  $\bY$ and $\bP$ are independent of the  total number of counts per sample.

The appropriate level $\alpha$ test of independence depends on the dimension and possible values of  $\bY$. For a univariate  $\bY$, the choice is among tests for equality of distributions if $\bY$ is categorical and among tests of independence between random variables if $\bY$ is continuous. For a  multivariate $\bY$, the choice is among tests of independence between a univariate random variable and a multivariate vector. For the null hypothesis is \eqref{eq:eqdist_latent2}, the choice is among tests of independence between two random vectors \citep{gretton2008kernel,szekely2009brownian,heller2012consistent}.
When the null hypothesis is \eqref{eq:eqdist_latent2} and $\bY$ is categorical, the popular PERMANOVA test \cite{anderson2001new} can be used on the vector of rarefied counts, see example in \S~\ref{sec:RDE_GUT}.

The test of $H_0^{\left(j\right)}$ makes no parametric assumptions on the distribution of $\bP$, or on the structural zeros. The assumption free test comes at a price of first having to rarefy $\bX_{i}(\be_j)$ to ${Z}_{i,\lambda_j}(\be_j)$. Normalization by rarefaction has been criticized since only part of the data is used for inference \citep{mcmurdie2014waste}. However, the alternative methods rely on parametric assumptions for modeling the data. Since little is known about the data generation mechanism, having no model assumptions is highly desired. Arguably, the potential power loss due to rarefaction is worth the gain in assurance that the correctness of discoveries does not hinge on model assumptions and sequencing resolution. We support our argument via examples and extensive simulations in \S~\ref{sec:simulations}-\S~\ref{sec:RDE_GUT}.

The above method depends on the particular rarefied sample that resulted in one draw. It may be tempting to consider several rarefied samples instead of relying on a single draw, but unfortunately averaging test statistics across multiple rarefactions of the data, or  averaging the rarefied draws themselves, will result in a non-valid method.  To see why, consider the case where the tested taxon $j$ is not differentially abundant, and the total number of counts available in the taxa with indices $\{j,b_1,\ldots,b_r\}$ for samples with group label $\bY=0$ is stochastically smaller than for samples with group label $\bY=1$, i.e., $\bb_j'\bX_i$ tends to be smaller if $\bY_i=0$ than if $\bY_i=1$. Hence, counts in samples with a trait of $\bY=0$ are more likely to be resampled across multiple rarefactions of the data compared to counts from samples with $\bY=1$. Therefore, the bivariate distribution of two rarefied draws taken from a single sample is different across different values of $\bY$. Specifically, multiple draws from a sample with $\bY=0$ will have a higher correlation compared to multiple draws from a sample with $\bY=1$.

Another approach we consider, since  the reference set of taxa has  a positive number of counts in all samples, is to reject the null hypothesis \eqref{eq:eqdist_latent} if the null hypothesis of independence between $\bX(\be_j)/\bb_j'\bX$ and $\bY$ is rejected using an appropriate level $\alpha$ test for the data $$\left \lbrace \left( \bX_i(\be_j)/\bb_j'\bX_i, \bY_i\right): i=1,\ldots,n \right \rbrace. $$
We will refer to this method of differential abundance testing as \emph{normalization by ratio}.
Our motivation for considering this test for differential abundance, is the fact that if the  null hypothesis \eqref{eq:eqdist_latent} is true, then the conditional expectation of $\bX(\be_j)/\bb_j'\bX$, given $\bP$, is independent of $\bY$, as follows from the result formally stated in the next proposition. 
\begin{prop}\label{prop-expectation}
	If $\bX\mid N, \bP\sim multinom(N, \bP)$, then
	$$E\left\lbrace\frac{\bX(\be_j)}{\max(1, \bb_j'\bX)}\mid \bP \right \rbrace = \frac{\bP(\be_j)}{\bb_j'\bP}Pr\left(\bb_j'\bX>0\mid \bP\right). $$
\end{prop}
See \S~\ref{app_A_proof_of_lemma} for the proof. 
The spread of $\bX_i(\be_j)/\bb_j'\bX_i$ may depend on $\bY$  when the  null hypothesis \eqref{eq:eqdist_latent} is true, so this approach can be approximately valid at best, but potentially more powerful when $\bb'\bX$  (i.e., the total count in the reference set is small) is small, than the valid test we suggested in steps 1-3 above.  We compare the two approaches for testing  \eqref{eq:eqdist_latent} in \S~\ref{sec:simulations}-\S~\ref{sec:RDE_GUT}.

The procedure for testing $H_0^{\left(\textbf{j}\right)}$ \eqref{eq:eqdist_latent2} is similar to the procedure for testing $H_0^{\left({j}\right)}$. For a non-negative integer $p$-vector $\textbf{v}$, let ${\bf U}\sim MHG\left(\lambda, \textbf{v}, M\right)$ denote the (multivariate hypergeomtric) distribution, so ${\bf U}$ is a random vector of dimension $p$ formed by counting the number of balls of types $1,...,p$, when sampling $\lambda$ balls, without replacement, from an urn containing $M$ balls out of which the number of balls of type $1,\ldots,p$ is $\bf v$.  Let $\lambda_\bj = \min_{i=1,\ldots,n} \bb_\bj'\bX_i$. For the $i$th observation, sample  ${\textbf{Z} }_{i, \lambda_\bj}(\be_\bj) \sim MHG\left(\lambda_\bj, \bX_i(\be_\bj), \bb_\bj'\bX_i\right).$ A test of independence between ${\textbf{Z} }_{ \lambda_\bj}(\be_\bj)$ and $\bY$ is a valid test for \eqref{eq:eqdist_latent2}, using the same reasoning as in the proof for Proposition \ref{prop-testvalidity}. A test of $H_0^{\left(\textbf{j}\right)}$ using \emph{normalization by ratio} will reject the null hypothesis using a level $\alpha$ multivariate test for the data:
$$ \left \lbrace \left( \bX_i(\be_\bj)/\bb_\bj'\bX_i, \bY_i\right): i=1,\ldots,n \right \rbrace. $$

\subsection{Choosing the reference taxa $(b_1,\ldots,b_r)$}\label{subsec:choosing_reference}
If the total absolute abundance of reference taxa is independent of the studied phenotype, rejections of \eqref{eq:eqdist_latent} (or \eqref{eq:eqdist_latent2}) could be interpreted as a change in the absolute abundance of the $j$th taxon (or to one or more of the taxa corresponding to the non-zero entries in $\bj$). If domain knowledge exists regarding taxa which are not associated with the condition examined, it can be used to construct a reference set of taxa. One possible technique to generate such a reference set is through a spike-in of synthesized DNA \citep[see Section "Spike-in log-ratio normalization" in ][]{quinn2018field} or bacteria not endemic to the ecosystem studied \citep{stammler2016adjusting}. Otherwise, when  the set of reference taxa to subsample against is not known a-priori, a data-adaptive method for finding the reference set is needed.

Without external information, we need to both identify the reference taxa, and then test with respect to this reference set, using the same dataset. If the absolute abundance of most taxa is independent of the studied phenotype, a large set of taxa whose inter-taxa proportion ratios are relatively stable could be taken as the reference set of taxa. It is important to identify the reference taxa without invalidating the testing that follows. If a large number of samples is available, the data could be split into two parts, the first part for reference selection, and the second part for testing. The reference selection procedure may include all taxa that appear least associated with the trait in the first part (as characterized by a large $p$-value for testing the independence of $\bX(\be_j)$ and $\bY$). However, if there is not enough data to spare for selection (since the testing of the second part will lack power), we suggest the following strategy.  Ideally, we would like the statistic used for taxa selection to be independent of the test statistic used for testing ${ Z}_{\lambda_j}(\be_j) \indep \textbf{Y}$ \citep{hommel2005tests}. As a first principle, our statistic for selection of reference taxa should not use the trait values.

Let 
$
SD_{j,k} = \stackrel[i=1]{n}{\mathrm{sd}}\left(log_{10}\left(\frac{\bX_i(\be_j)+1}{\bX_i(\be_k)+1}\right)\right),
$
where $sd$ is the sample standard deviation taken over $n$ values. The statistic for selection of reference taxa is the median, 
$
S_j =\mathrm{median}_{\{k: k\neq j, k=1,\ldots, m\}}\left(SD_{j,k}\right).
$
The resulting reference set is $B = \{j|S_j \leq S_{crit}\}.$
The appropriate value of $S_{crit}$ may be application specific, see \S~\ref{app:ref_score_distribution} for details. 

We require that the total number of counts in the reference set be positive for all samples, to ensures that for each taxon tested $\lambda_j>0$, but the total number need not be very high (the greater the reference set,  the greater  the risk that it includes differentially abundant taxa). Therefore, following selection of the potential reference set, we proceed to add or remove reference taxa, depending on whether the minimal number of total reference counts per sample is too small or too large. 
If it is too small, e.g., less than 10, we increase $S_{crit}$ until the minimum of 10 total reference counts is reached by all samples. If it is too large, e.g., more than 200, $S_{crit}$ is reduced accordingly. 

\section{A simulation study}\label{sec:simulations}

A simulation study was performed to compare the power and error rate control of various tests for discovering the differentially abundant taxa. For simplicity, we focus on settings where $\bY$ is a binary variable indicating group membership.  

The newly suggested procedures for \underline{d}ifferential  \underline{a}bundance testing with \underline{comp}ositionality adjustment are denoted by DACOMP, DACOMP-t, and DACOMP-ratio.
These tests use a reference set that is adaptively chosen from the data as described in \S~\ref{subsec:choosing_reference}, with $S_{crit} = 1.3$. 
The chance that differentially abundant taxa erroneously enter the reference set was negligible in the vast majority of our simulated settings,  see \S~\ref{app:ref_score_distribution} for details and \S~\ref{app:naive reference selection} for alternative reference selection methods. 
DACOMP and DACOMP-t follow the procedure in \S~\ref{subsec:oracle_test}, and they 
differ  only with regard to the two-sample test carried out in Step 3: Wilcoxon rank-sum test for DACOMP, and Welch two sample  t-test  on transformed counts,   $log\left({Z}_{i,\lambda_j}(\be_j)+1\right)$, for DACOMP-t. DACOMP-ratio follows the \emph{normalization by ratio} approach in \S~\ref{subsec:oracle_test}, with Wilcoxon rank-sum test as the two-sample test. 

Previously suggested tests considered are: ANCOM  \citep{mandal2015analysis}, as implemented in version 1.1-3 of the \emph{ANCOM} package; W-FLOW, Wilcoxon rank sum tests with the correction by \citet{vandeputte2017quantitative}; W-CSS and W-TSS,  Wilcoxon rank sum tests with the CSS and TSS normalization, respectively,  with  W-CSS as implemented in the software package \emph{metaGenomeSeq} in R \citep{metaGenomeSeq_Package} in version 1.24-1; ALDEx2-t and ALDEx2-W \citep{fernandes2013anova}, using the two-sample Welch t-test and Wilcoxon rank-sum test, respectively, as implemented in version 1.16-0 of the \emph{ALDEx2} package ; WRENCH \citep{kumar2018analysis}, implemented in version 1.2-0 of the \emph{wrench} package, with default parameters (it makes use of the tests of differential abundance implemented in the `deseq2` software package \citep{love2014moderated});  HG, Fisher's exact test against a reference set. The reference set for  HG was the oracle set that includes all non differentially abundant taxa with $S_{crit} = 1.3$, in order to demonstrate that the test is biased due to a failure to account for over dispersion (rather than due to the reference set being contaminated with signal).

For error control, we chose the false discovery rate (FDR, \citealt{benjamini1995controlling}). ANCOM carries out its own multiplicity correction aimed at FDR control.  For all other methods, we applied the Benjamini-Hochberg (BH)  procedure \citep{benjamini1995controlling} at level $q=0.1$. We chose the BH procedure since  empirical evidence and simulations suggest it controls the FDR for most dependencies encountered in practice, including microbiome applications \citep{jiang2017discrete}, even though  the theoretical guarantee is only for independence or a type of  positive dependence. The family of tests is smaller for the new DACOMP tests that for the other tests, since the taxa in the reference set are not tested for differential abundance. 

We  considered settings with overdispersion and compositionality, by  resampling from a microbiome dataset,
and we display the results 
for ANCOM, W-FLOW, W-CSS, DACOMP, DACOMP-ratio, ALDEx2-t and HG, which represent key approaches. The other   results are detailed in \S~\ref{app:additional sim results}. We  considered in \S~\ref{app:additional sim results}  
also the following  additional settings: 
a setting where  sequencing depth varies across groups (as discussed, e.g.,  in \citealt {silverman2018statistical}), for which we show that
only  DACOMP and DACOMP-t provides adequate control over false positives; a (less realistic) setting   where the total microbial load of the differentially abundant taxa is identical across study groups so  marginal methods provide a valid method of testing since there is no bias due to compositionality, for which we show that the loss of power when using DACOMP  is small; and a setting where only the rare taxa are differentially abundant, causing a severe inflation of  false positives  for some competitor methods.
The simulation results are based on 100 replications.

\subsection{Data generation}\label{subsec:sim_gut}

The data used as a basis for this simulation is described in \citet{vandeputte2017quantitative}, as the 'Disease cohort' of the study. The V4 region of the 16S gene was amplified and sequenced from fecal samples of 66 healthy subjects. In addition, the number of bacteria per gram were measured using a flow cytometer. 
We picked sOTUs ( a type of ASVs, see \S~\ref{sec:Introduction})  using the method of  \citet{amir2017deblur}. sOTU length was set to the default value of 150 base pairs. In total, $1722$ sOTUs were selected. All sOTUs which appeared in less than 4 subjects were removed from the data, leaving $m=1066$ sOTUs. The median number of reads across subjects was $ N_{reads} = 22449$ reads across the $1066$ sOTUs.

For a simulated dataset, a total of 60 'healthy' and 60 'sick' subjects were sampled. The vector of counts for the 'healthy' $i$th was generated by the following steps: (1) the 16S vector of counts and a flow cytometric measurement, denoted by $\textbf{u}_i^H$ and $C_i^{H,flow}$, were recorded for a randomly selected subject, so ${\boldsymbol \mu}_i^H = C_i^{H,flow}\times \textbf{u}_i^H /\textbf{1}'\textbf{u}_i^H$ is the  unobserved  abundance vector of taxa; (2) the total number of reads, $N^H_i$, was sampled from the Poisson distribution with parameter $N_{reads}$; (3)  $\bf{X}_i$ was sampled from $multinomial(N_i^H,\bP_i)$, where $\textbf{P}_i ={\boldsymbol \mu}_i^H/\textbf{1}'{\boldsymbol \mu}_i^H$.

The vector of counts for the 'sick' subjects were generated in a manner similar to steps 1-3 above, with the following changes in $m_1\in\{10,100\}$ differentially abundant taxa selected at random. For the $i$th 'sick' subject,  each taxon $j$ associated with the disease had a chance of 0.5 to experience an increase in its absolute abundance of bacteria in each 'sick' subject. The random number of bacteria added to the absolute abundance of the $j$th taxon was sampled, independently for each entry, from $N\left(\mu_{i,j},\mu_{i,j} \right)$ 
, where $\mu_{i,j} = \lambda_{effect}\times C_i^{S,flow}\times \delta_j/m_1$. The parameter 
$\lambda_{effect}$ dictates the expected increase in the host microbial load due to the simulated conditioned , e.g., $\lambda_{effect} = 1.0$ indicates an expected increase of 100\% in the total host microbial load. The parameter $\delta_j$ sets the strength of association of a specific taxon with the simulated condition. We considered the range of values $\lambda_{effect}=0,0.5,1.0,...,3.0$ and $\delta_j=\{0.5,1.0,1.5\}$. Clearly,  the resulting   abundance vector of taxa, ${\boldsymbol \mu}_i^S$, differs in distribution from   ${\boldsymbol \mu}_i^H$ only in the $m_1$ coordinates where counts were added, and only for these  coordinates  the null hypothesis \eqref{eq:eqdist_latent} is false.  

\subsection{Results}\label{subsec:sim_gut_results}

\begin{figure}[htbp]
	\centering
	\includegraphics[width=1.0\textwidth]{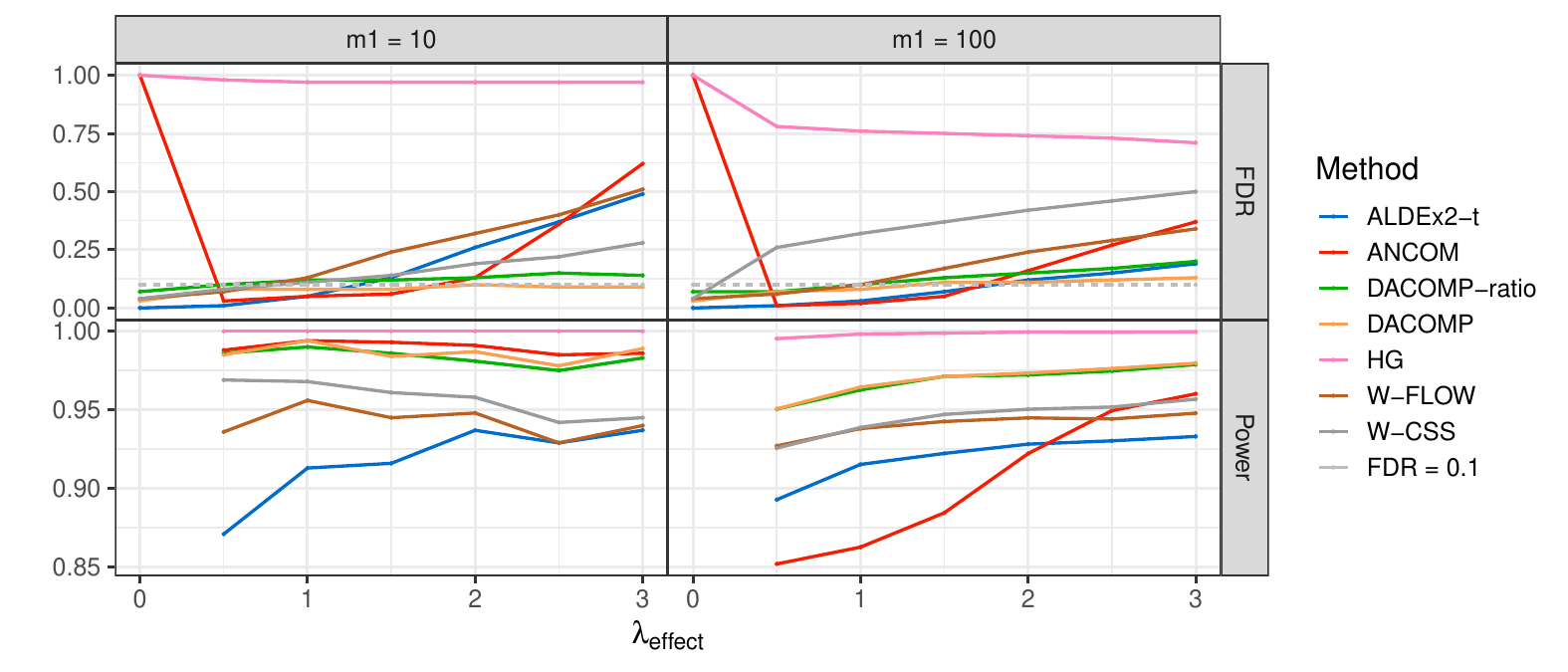}
	\caption{Estimated FDR and power versus $\lambda_{effect}$ for DACOMP and competitors in the simulation settings of \S~\ref{subsec:sim_gut}.  The level of the BH procedure was  $q=0.1$ (in dashed gray).  The maximal standard error for FDR and power was 0.04 and 0.73, respectively.}
	\label{figure:sim_p1_TP_and_FDR}
\end{figure}

Figure~\ref{figure:sim_p1_TP_and_FDR} shows the estimated FDR and power for each method, for the different scenarios. DACOMP is the only method controlling FDR across all scenarios considered. For the global null setting ($\lambda_{effect}=0$), only ANCOM and HG do not control the FDR. For HG this is expected since we have overdispersion in the data. For ANCOM, we have observed that generally, under the global null, FDR is not controlled. 
In \S~\ref{app:ANCOM_global_null}, we present additional scenarios with no differentially abundant taxa where ANCOM does not control the FDR.
ANCOM and W-FLOW lack FDR control when  $\lambda_{effect} \ge 2.0$. For ANCOM, this could be attributed either to the empirical decision rule being invalid or to mistreatment of technical zeros by using a pseudocount. For W-FLOW, the lack of FDR control can be attributed to mistreating technical zeros as well: W-FLOW uses a multiplicative factor to correct for compositional bias, providing no solution for technical zeros. ALDEx2-t provides FDR control for $m_1=100$ but not for $m_1=10$. For DACOMP-ratio, the inflation is largest with $\lambda_{effect}=3$, with a maximum realized FDR value of  0.17. 

For $m_1=10$ the power is close to one for all methods. For $m_1=100$, DACOMP has the highest statistical power, despite being the only valid procedure. The increase in power results mainly from excluding the reference set of taxa from testing: the mean size of selected reference sets across scenarios varied from 506 for $m_1=100$ and $\lambda_{effect}=0.5$,  to 691 for $m_1=10$ and  $\lambda_{effect}=3.0$ (the standard error was $<15$).  While DACOMP has the highest expected number of true discoveries, its expected number of discoveries is substantially lower, as other methods do not provide adequate FDR control. For example, for the case where  $\lambda_{effect}=2.5$ and $ m_1 = 100$, W-CSS has 176 discoveries on average, but only 95 true discoveries.

\section{A study of  Crohn's disease}\label{sec:RDE_GUT}  \citet{vandeputte2017quantitative} examined  fecal samples 29 subjects with  Crohn's Disease (CD) and 66 healthy controls. 
All subjects had 16S profiling for their fecal samples taken along with a microbial load count, given in number of bacteria per gram of fecal material. 

The flow-cytometry measurements showed that the total abundance of the microbiota is much lower for subjects with CD: the median microbial load  was  $3.76\cdot 10^{10}$ and $1.16\cdot10^{11}$  bacteria per gram for subjects with CD and for healthy subjects, respectively. 
With such a high change in microbial load, it is implausible to assume most taxa have not altered their absolute abundance across study groups in the presence of CD. However, the majority of taxa may still be non differentially abundant. 

DACOMP is particularly suitable for  this study, since  the total number of reads is associated with CD:  the median number of 16S reads across subjects with and without CD was $18874$ and $22871$, respectively.
Our units for analysis were the 1569 sOTUs that appeared in at least in two subjects.

\subsection{A univariate analysis}\label{RDE_GUT_UNIVARIATE}
Table~\ref{table:RDE_GUT_Shared} shows the number of taxa discovered for each method, along with the number of discoveries shared by the different methods. Procedures DACOMP and ALDEx-t have a similar number of discoveries, which is 
substantially lower than with ANCOM, W-FLOW and W-CSS. 
W-FLOW uses an additional flow-cytometric measurement, yet it has a lower number of discoveries than ANCOM and W-CSS.
The difference in discoveries between DACOMP and DACOMP-ratio may from either a reduction in power due to subsampling step done in DACOMP, or DACOMP-ratio not controlling the rate of false positive discoveries.
Out of the 1569 taxa, 1305 had less than 10 counts on average. For these relatively rare taxa, the methods show little agreement.   ANCOM, W-CSS, Aldex2-t, and DACOMP discover 44, 102, 0, and 6 taxa, respectively, in addition to the ones discovered by W-Flow, suggesting that ANCOM and W-CSS may have a non-negligible number of false discoveries. For the remaining 264 taxa, the agreement between  methods is good, see Figures \ref{fig:Crohn_shared_Venn_Abundant} and \ref{fig:Crohn_shared_Venn_Rare}  in \S~\ref{app:additional_CD_results}. 


\begin{table}[ht]
	\caption{Number of discoveries by each method, for the data of \citet{vandeputte2017quantitative}. The number of discoveries by each method on the diagonal, and shared with the other methods on the off-diagonal entries. 
		For DACOMP and DACOMP-ratio $S_{crit}=1.3$.}
	\label{table:RDE_GUT_Shared}
	\small
	\begin{tabular}{lrrrrrrr}
		\hline
		Method  & ANCOM & W-FLOW & W-CSS & ALDEx2-t & DACOMP & DACOMP-ratio \\ 
		\hline
		ANCOM & 216 & 154 & 189 & 103 & 101 & 123 \\ 
		W-FLOW &  & 195 & 149 & 101 & 105 & 121 \\ 
		W-CSS &  &  & 276  & 95 & 104 &  132 \\ 
		ALDEx2-t &  &  &  & 103 &  85 &  94 \\ 
		DACOMP & &  &  &  & 123 & 113 \\ 
		DACOMP-ratio & &  &  &  &  & 163 \\ 
		\hline
	\end{tabular}
\end{table}


\subsection{A multivariate analysis}\label{RDE_GUT_MULTIVARIATE}
In order to identify the genera which are differentially abundant, 
the sOTUs  were assigned taxonomy level data using a taxonomy classifier \citep{wang2007naive} as implemented in the \verb|assignTaxonomy| function of the \verb|dada2| \citep{callahan2016dada2} software package. The classifier used the Green Genes taxonomic training set \citep[version 13.8, ][]{desantis2006greengenes} as a reference database. 
Using 80 bootstrap permutations,  871 sOTUs were assigned to 62 genera that contained more than a single sOTU,
with a median genus size of 5 sOTUs.

For a specific genus
, let $\bf e_{\bf j}$ and $\bf b_{\bf j}$ denote the binary vectors of length 1569 with either 'ones' in the vector entries corresponding to genus $g$ alone, or to genus $g$ and the reference taxa, respectively. We tested the null hypothesis \eqref{eq:eqdist_latent2} by applying the PERMANOVA test in order to discover whether  the rarefied counts, $\textbf{Z}_{i, \lambda_\bj}(\be_\bj)$, for the DACOMP approach, or   $\bX_i(\be_\bj) / \bb_\bj'\bX_i$, for the DACOMP-ratio approach, are associated with CD status.  Our metric was the robust  Mahalanobis distance detailed in Chapter 8.3 of \cite{rosenbaum2010design}), which protects against  outliers and takes the correlation among counts into account.   
We also tested \eqref{eq:eqdist_latent} (as in \S~\ref{RDE_GUT_UNIVARIATE}) by treating each genus as a taxon, where the observed taxon count is the sum of  sOTU counts in the  genus. The family of 62 genera were tested using  the BH procedure  at level $0.1$. 
Table~\ref{table:RDE_GUT_MULTIVARIATE} shows the number of discoveries by each method, as well as the overlap across methods. Interestingly, for each normalization approach, about a third of the genera discovered by the multivariate test statistic are not discovered by the univariate test statistic (and vice versa).   

\begin{table}[ht]
	\caption{Number of genera discovered (out of 62) as differentially abundant  using the PERMANOVA test in the DACOMP approach (Multi) and the DACOMP-ratio approach (Multi-ratio), and using the  Wilcoxon test at the genera  level in the DACOMP approach (Uni) and the DACOMP-ratio approach (Uni-ratio).  The number of discoveries by each method on the diagonal, and shared with the other methods on the off-diagonal entries.  
	}
	
	\label{table:RDE_GUT_MULTIVARIATE}
	\centering
	\begin{tabular}{rrrrr}
		\hline
		DACOMP: & Multi & Multi-ratio & Uni & Uni-ratio \\ 
		\hline
		Multi &  18 &  17 &  12 &  14 \\ 
		Multi-ratio &  &  32 &  17 &  20 \\ 
		Uni &  &  &  23 &  22 \\ 
		Uni-ratio &  &  &  &  33 \\ 
		\hline
	\end{tabular}
\end{table}

\section{Final remarks}\label{sec:Discussion}
In this paper, we provide a novel method for discovering differentially abundant taxa with minimal assumptions. We demonstrated the validity of our method, DACOMP,  and the potential inflation of false positives of other methods. We also showed the good power properties DACOMP. The novelty of our approach lies in  replacing the common practice of normalizing count vectors by a comparison of the taxa of interest with a reference set of taxa, after rarefying the counts so that the rarefied counts of non-differentially abundant taxa are independent of the trait. In settings where the total number of counts in the reference set is small, we suggested DACOMP-ratio, which may be biased but avoids the rarefying step that may hinder power. In numerical comparisons, we showed that with DACOMP-ratio we can gain power but at a price of an inflation in the type 1 error probability. However, this inflation is typically small in comparison with the inflation incurred by other methods.  

We provided empirical evidence that our approach is useful  in a study of Chron's disease, where the compositional effect is large. 
In addition, we analyze in \S~\ref{app:RDE_HMP}  the differential abundance of taxa across adjacent body sites in the human body using data from the Human Microbiome Project \citep{gevers2012human}, where DACOMP discovers a considerable number of taxa as differentially abundant. 
In \S~\ref{app:RDE_DILUTION}, we analyze data from a stool sample dilution experiment \citep{stammler2016adjusting}, where fecal samples were first diluted at different ratios, and then 'spiked-in' with a known load of three types of bacteria. Unlike previous examples, for this data set, the "ground truth" for differential abundance is known. Moreover, the traits examined are continuous: the dilution factor and  the microbial load spiked-in. Therefore, we tested \eqref{eq:eqdist_latent} using Spearman's correlation test, and we showed that DACOMP detects the true differentially abundant taxon, and that some of the other methods have an inflation of false positives.  

A crucial step in our approach is the identification of an appropriate reference set. In  \S~\ref{subsec:choosing_reference} we provided a data adaptive method, which 
avoids using the trait values explicitly for reference selection. 
However, the reference selection statistics, $S_j$, are not  independent of the trait $\bY$ if the global null is false, since for two non differentially abundant taxa 
$\textbf{P}(\textbf{e}_j)/ \textbf{P}(\textbf{e}_k)$ is independent of the measured trait, but 
$(\textbf{X}(\textbf{e}_j) +1)/(\textbf{X}(\textbf{e}_k)+1)$ may not be. 
In our experiments with a small number of samples, we demonstrated empirically that the selection does not  invalidate the testing procedure. For large enough sample sizes, the data can be randomly split into two parts, with the first group used for reference selection and the second group used for testing, ensuring the statistics used for reference selection are independent of the test statistics. We leave for future research the goal of designing methods for reference set selection that are theoretically valid yet more efficient than sample splitting.


Other fields of study that gather data by sequencing PCR amplicons also make use of statistical methods aimed at analyzing compositional data, for example: RNA-seq \citep{quinn2018field} , metabolomics \citep{kalivodova2015pls}, and shotgun sequencing techniques for microbiome data \citep{calle2019statistical}. Adapting DACOMP and DACOMP-ratio  to such datasets is an interesting direction for future work. 

\appendix
\section{Proofs}\label{app_A_proof_of_lemma}
\noindent{\bf Proof of Proposition \ref{prop-testvalidity}}.\\
Since $\lambda_j$ is a function of the total counts of taxon $j$ and the taxa in the reference set, the proof follows if the rarefied counts, conditional on these total counts, depend only on $\lambda_j$ and $\bP(\be_j)/\bb_j'\bP_i(\bb_j)$. It is straightforward to show that this is indeed the case, using the following lemma.

\begin{lemma}\label{lemma:binomial}
	Let $\left(U,V,W\right)\sim multinom\left(N,\left(P_U,P_V,1-P_U-P_V\right)\right)$ and $\tilde{U}|U,V,\lambda\sim hypergeom\left(\lambda,U,U+V\right),$  then:
	$$\tilde{U}|\lambda,P_U,P_V, U+V \sim bin\left(\lambda,\frac{P_U}{P_U+P_V}\right), $$
	where $hypergeom\left(t,z,z+w\right)$ is the distribution of the number of special items sampled when selecting $t$ distinct items from a population of $z+w$ items, $z$ of which are special.
\end{lemma}
\begin{proof}
	It is easy to see that $U|\{U+V=a\}\sim Bin\left(a,{p_U}/{\left(p_U+p_V\right)}\right)$. The value of $P(\tilde{U} = x|\lambda_j,P_U,P_V,U+V=a)$ can be computed from the law of total probability, summing over the possible values of $U$:
	
	$$P(\tilde{U} = x|\lambda_j,P_U,P_V,U+V=a) = \sum_{b=x}^{a-\lambda+x}P\left(\tilde{U}=x|U=b,U+V=a,\lambda\right)\times $$ $$P\left(U=b|U+V = a\right) = $$
	
	$$\sum_{b=x}^{a-\lambda+x}\frac{{b \choose x}{a-b \choose \lambda-x}}{{a \choose \lambda}}{a \choose b}\left(\frac{P_U}{P_U+P_V}\right)^{b}\left(\frac{P_V}{P_U+P_V}\right)^{a-b},$$
	where we used the fact that $b$ is between $x$ (all items of category $U$ were sampled in $\tilde{U}$) and $a-\lambda+x$ (all items of category $V$ were sampled). Expanding all combinatorial factors, and substituting the index variable to $c=b-x$, the former expression can be written as:
	$$\sum_{c=0}^{a-\lambda}
	{\lambda \choose x}{a-\lambda \choose c}\left(\frac{P_U}{P_U+P_V}\right)^{x +c}\left(\frac{P_V}{P_U+P_V}\right)^{(\lambda-x)+(a-\lambda-c)}.$$
	
	We recognize that the index variable $c$ sums over a binomial distribution probability function, simplifying the expression to
	$
	{\lambda\choose x}\left(\frac{P_U}{P_V+P_U}\right)^{x}\left(\frac{P_V}{P_U+P_V}\right)^{(\lambda-x)},
	$
	as required.
\end{proof}

\noindent{\bf Proof of proposition \ref{prop-expectation}}.
\begin{proof}
	\begin{eqnarray}
	&& E\left\lbrace\frac{\bX(\be_j)}{\max(1, \bb_j'\bX)} \mid \bP \right \rbrace = 
	E\left\lbrace\frac{\bX(\be_j)}{\bb_j'\bX} \mid \bP, \bb_j'\bX>0 \right \rbrace Pr\left(\bb_j'\bX>0\mid \bP\right) \nonumber \\
	&& = E\left [\frac{1}{\bb_j'\bX}E \left\lbrace\bX(\be_j) \mid \bP, \bb_j'\bX \right \rbrace \mid  \bP, \bb_j'\bX>0 \right] Pr\left(\bb_j'\bX>0\mid \bP\right) \nonumber \\
	&& = \frac{\bP(\be_j)}{\bb_j'\bP}Pr\left(\bb_j'\bX>0\mid \bP\right), \nonumber 
	\end{eqnarray}
	where the last equality follows since $\bX(\be_j) \mid \bP, \bb_j'\bX$ is binomial with parameters $\bb_j'\bX$ and  $\frac{\bP(\be_j)}{\bb_j'\bP}$ and thus with expectation  $\bb_j'\bX\frac{\bP(\be_j)}{\bb_j'\bP}$. 
	
\end{proof}

\section{Supplementary Material}
The methods presented in this paper for differential abundance testing and reference selection are available as an R package on Github (\href{https://github.com/barakbri/dacomp}{github.com/barakbri/dacomp}). Source code and instructions describing how to reproduce the results in this paper are found on (\href{https://github.com/barakbri/CompositionalAnalysis_CodeBase}{github.com/barakbri/CompositionalAnalysis\_CodeBase}).

An additional PDF file with supplementary material contains the following Sections:\S~\ref{sm-lambdapos-example} exemplifies why excluding samples based on the counts available under the reference set of taxa may induce bias in testing; 
\S~\ref{app:additional sim results} contains additional simulation results for 
additional scenarios and competitor methods; \S~\ref{app:additional_CD_results} presents additional results for Crohn's disease data analysis example ;\S~\ref{app:appendix_B} contains further examination of the reference selection procedure, discusses how $S_{crit}$ was set, and reviews alternative reference selection procedures; \S~\ref{app:ANCOM_global_null} contains a simulation analyzing the control of false positive discoveries by ANCOM when $m_1 = 0$; \S~\ref{app:RDE_HMP} describes an analysis of differential abundance in the Human Microbiome Project (HMP) across pairs of body sites in the human body;finally, \S~\ref{app:RDE_DILUTION} presents an analysis of differential abundance with respect to a continuous trait, using a test based on the Spearman rank-correlation.

\bibliography{BRILL}

\renewcommand{\thesection}{S\arabic{section}}
\renewcommand{\thesubsection}{S\arabic{section}.\arabic{subsection}}

\newpage
\setcounter{section}{0}

\section{An example of how excluding samples based on reference size may induce bias}\label{sm-lambdapos-example}
\begin{figure}
	
	\centering
	\includegraphics[width=1.0\textwidth]{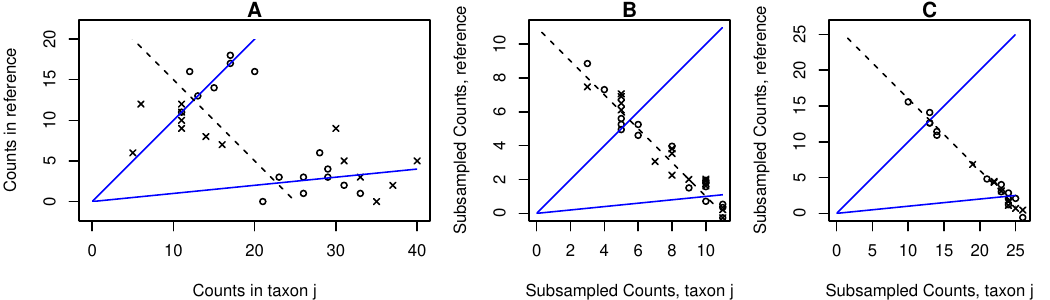}
	\caption{ \textbf{(A)}\
		The counts in taxon $j$ vs. total counts in reference taxa. Crosses and circles represent samples with $\textbf{Y}=0$ and $\textbf{Y}=1$, respectively. Blue lines form the two possible values of  ${\textbf{P}(\textbf{e}_j)}/{\textbf{b}_j}'\textbf{P}$, a ratio of $1:1$ or $1:10$.
		The dashed black line represents a total of 25 counts observed in taxon $j$ and the reference set altogether. \textbf{(B)} Observations are subsampled to highest possible depth without removing samples, $\lambda_j$ is shown by the black dashed line. To account for ties in the data, coordinates for the vertical axis were jittered.  \textbf{(C)} Observations with less than $25$ counts in taxa with indices $j\cup B$ were removed. The remaining observations, above the dashed line in subplot A, were subsampled to depth $25$ and depicted in graph C.
	}
	\label{figure:NO_FILTER_GRAPH}
\end{figure}

\emph{Example 3: excluding samples based on reference size may induce bias.} Consider the setting where a non differentially abundant taxon $j$ is tested for differential abundance against a binary valued trait, using a reference set of taxa, denoted by the binary vector $\b b$. Let $\textbf{b}_j = \textbf{b}+\textbf{e}_j$. We assume ${\textbf{P}(\textbf{e}_j)}/{\textbf{b}_j}'\textbf{P}$ obtains the values $0.5$ and $0.9$ with equal probability.  We observe a random sample, $n= 32$, with $\textbf{Y}_i=0$ for $i=1,...,16$ and $\textbf{Y}_i=1$ for $i=17,...,32$. Due to a change in the absolute abundance of the differentially abundant taxa, the number of counts available under taxa $j$ and the reference set differs for different value of $\b Y$. The total number of observed counts in taxa $j$ and the reference set for samples $i\in\{17,...,32\}$ is distributed $Pois\left(30\right)$. For samples $i\in\{1,...,16\}$, the total number of counts observed in these taxa depends on ${ \textbf{P}(\textbf{e}_j)}/{\textbf{b}_j}'\textbf{P}$: it is distributed $Pois\left(20\right)$ if ${ \textbf{P}(\textbf{e}_j)}/{\textbf{b}_j}'\textbf{P} = 0.5$, and $ Pois\left(40\right)$ otherwise.  Figure~\ref{figure:NO_FILTER_GRAPH} shows that by subsampling to the minimum depth without exclusion of samples, the resulting samples appear to come from the same distribution, as expected (subplot B). However, if subsampling to a depth that requires samples below that depth to be excluded, the resulting samples no longer appear to come from the same distribution, potentially leading to spurious discovery claims (subplot C).

\section{Additional simulation results}\label{app:additional sim results}
This section of the supplementary material describes simulations results for additional settings and methods. In \S~\ref{app:additional_sim results_varying_seq_depth} we present simulations results for settings where sample sequencing depth of is confounded with the trait of interest. We show only DACOMP provides valid statistical inference in terms of type I error, since it conditions on the number of reads observed. In \S~\ref{subsec:sim_multinom} we discuss settings where only the rare taxa are differentially abundant, causing a severe inflation of the false positive rate for some competitor methods. In  \S~\ref{subsec:sim_nocomp} we discuss settings where the total microbial load of the differentially abundant taxa is identical across study groups and marginal methods provide a valid method of testing. We show that the loss of power when using DACOMP compared to methods aimed at inference on the change of  marginal distributions alone is small, for cases where compositional bias is not an issue. The inferential methods presented in the former two sections are identical to the ones presented in \S~\ref{subsec:sim_gut}. Finally, in \S~\ref{app:additional competitors} we discuss the simulation results for additional competitors methods, for the settings presented in\S~\ref{subsec:sim_nocomp}, \S~\ref{subsec:sim_gut}, and \S~\ref{subsec:sim_multinom}.

\subsection{Additional simulations where the sequencing depth varies between study groups}\label{app:additional_sim results_varying_seq_depth}
This subsection discusses simulation settings where sequencing depth of different samples differs between study groups. A confounding effect of sequencing depth may be observed in real data due to a difficulty to extract DNA that arises in only some of the studied sample groups \citep[see discussion of systemetic biases in ][]{silverman2018statistical}. It is interesting to assess the power and control of false positive discoveries of the methods compared in the paper under such biases. We show that DACOMP and DACOMP-t, which employ the modified rarefaction technique presented in \S~\ref{subsec:oracle_test} are the only methods that provide control of type I error when sample sequencing depth depends on the group labeling of observations.

For the simulations discussed in this subsection, data was generated similar to \S~\ref{subsec:sim_gut}, with the following difference. For the settings discussed in \S~\ref{subsec:sim_gut}, the number of sequenced reads for samples taken from healthy and sick subjects, denoted by $N_i^{H}$ and $N_i^S$,respectively, where sampled from $Pois\left(N_{reads}\right)$, where  $N_{reads} \equiv 22449$. For the settings in this subsection, data was either generated with $N_i^S\sim Pois\left(3\cdot N_{reads}\right)$ or $N_i^S\sim Pois\left(\frac{1}{3}\cdot N_{reads}\right)$. We will refer to these settings as "Group S Oversampled" and "Group S Undersampled", respectively.  Simulations consisted of $100$ simulated datasets for each value of $\lambda \in \{0,1,2,3\}$ and $m_1\in\{10,100\}$. The setting $\lambda=0$ is the global null case, with no differentially abundant taxa.

\begin{figure}[htbp]
	
	\centering
	\includegraphics[width=1.0\textwidth]{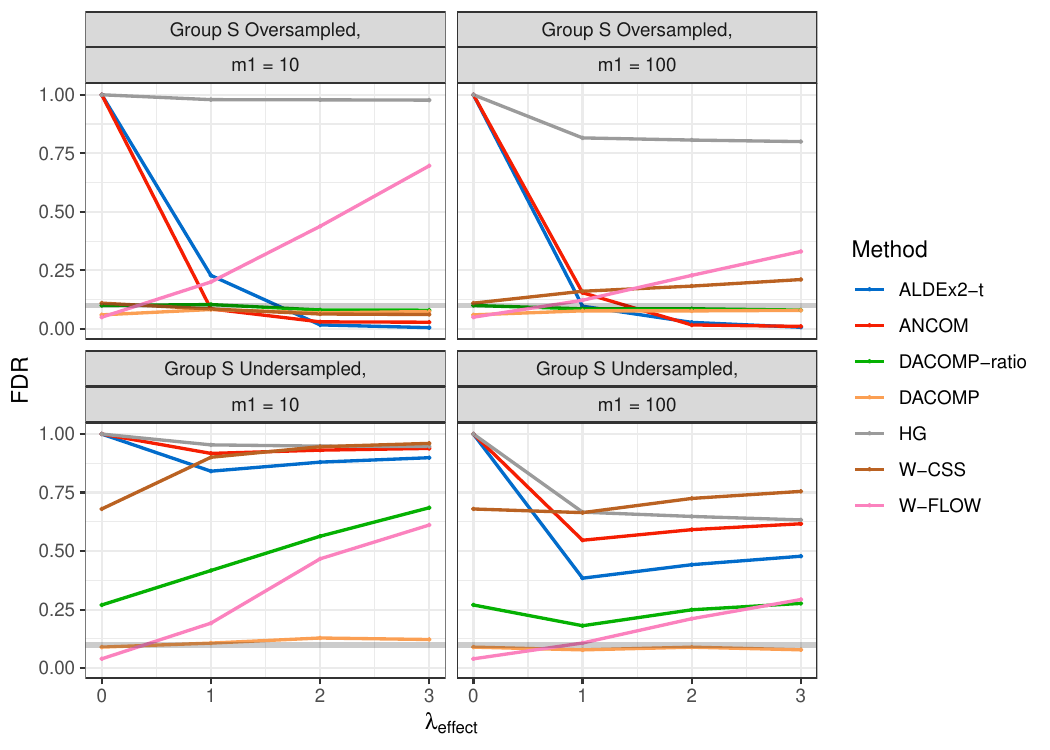}
	\caption{Estimated FDR of DACOMP and competitors for the simulation settings of \S~\ref{app:additional_sim results_varying_seq_depth}. The Y axis represents estimated FDR, the X axis represents $\lambda_{effect}$, the increase in percents in the microbial load of a sample with the simulated condition, e.g. a value of $1.0$ means a $100\%$ increase in the microbial load. The maximal standard error is 0.05. BH procedure was used at $q=0.1$. The gray line marks the value $q=0.1$.
	}
	\label{figure:app_sim_p4_FDR_additional_1}
\end{figure}

\begin{figure}[htbp]
	
	\centering
	\includegraphics[width=1.0\textwidth]{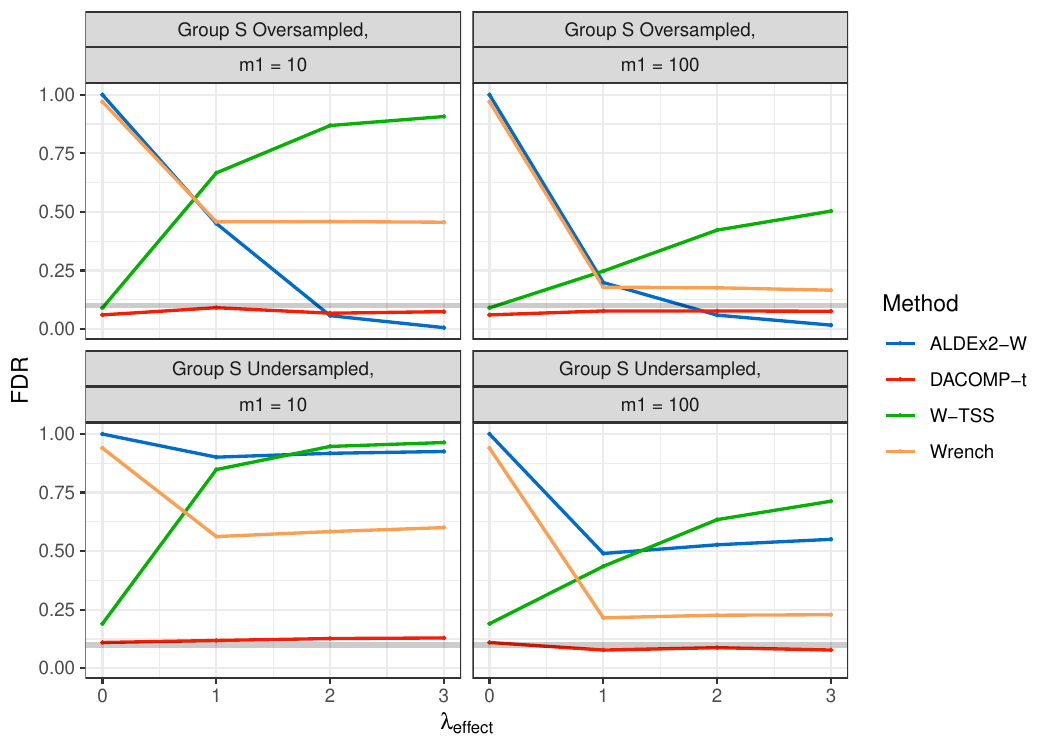}
	\caption{Estimated FDR of DACOMP and competitors for the simulation settings of \S~\ref{app:additional_sim results_varying_seq_depth}. The Y axis represents estimated FDR, the X axis represents $\lambda_{effect}$, the increase in percents in the microbial load of a sample with the simulated condition, e.g. a value of $1.0$ means a $100\%$ increase in the microbial load. The maximal standard error is 0.05. BH procedure was used at $q=0.1$. The gray line marks the value $q=0.1$.
	}
	\label{figure:app_sim_p4_FDR_additional_2}
\end{figure}

Figures \ref{figure:app_sim_p4_FDR_additional_1}-\ref{figure:app_sim_p4_FDR_additional_2} describe the average FDR of DACOMP and competitors, in a format similar to the graphs of \S~\ref{sec:simulations} and \ref{app:additional sim results}: Figure~\ref{figure:app_sim_p4_FDR_additional_1} compares DACOMP to the methods presented in \S~\ref{sec:simulations}, while Figure~\ref{figure:app_sim_p4_FDR_additional_2} compares DACOMP-t to the additional methods discussed in the beginning of this section.
For $m_1=10$ with "Group S Oversampled", all methods but HG and W-FLOW maintain control of the false positives rate and FDR. For ANCOM, W-CSS and ALDEx-2, the inflation of type I error that was observed in the corresponding setting in \S~\ref{sec:simulations}, Figure~\ref{figure:sim_p1_TP_and_FDR} is not observed under the setting with  $m_1=10$ and with "Group S Oversampled". The retained control of type I error is due to the distribution of non differentially abundant taxa in group S being less discrete and containing less technical zeros compared to the setting where the sequencing depth of samples is equally distributed between study groups. For the setting "Group S Oversampled" with $m_1=10$, W-CSS fails to provide control of false positive discoveries, in addition to HG and W-FLOW.

For $m_1\in {10,100}$ with "Group S Sndersampled", the effect described above is reversed: the counts distribution of non differentially abundant taxa in samples from group S becomes even more discrete compared to the cases of \S\ref{sec:simulations}, Figure~\ref{figure:sim_p1_TP_and_FDR}, as less counts are observed in samples from group S to begin with. For settings where "Group S is Undersampled", DACOMP and DACOMP-t alone provide control of the false postive rate and FDR. For DACOMP-ratio, which demonstrated a maximal FDR of $0.17$ in \S\ref{sec:simulations}, we now observe a maximal FDR of $0.28$ for $m_1=10$ and $0.69$ for $m_1=100$.

\begin{figure}[htbp]
	\centering
	\includegraphics[width=1.0\textwidth]{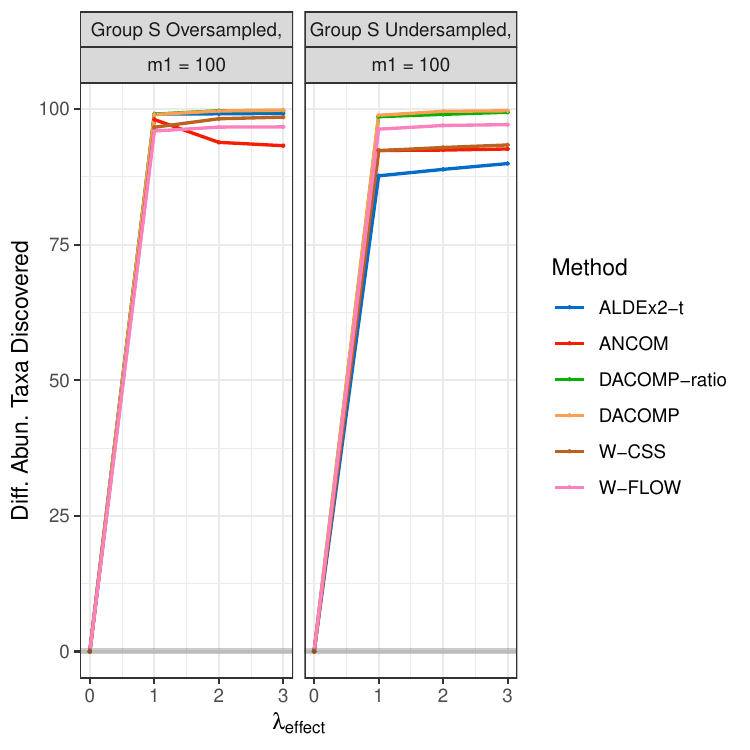}
	\caption{Estimated power of DACOMP and competitors for the simulation settings of \S~\ref{app:additional_sim results_varying_seq_depth}. The Y axis represents average number of true discoveries, the X axis represents $\lambda_{effect}$, the increase in percents in the microbial load of a sample with the simulated condition, e.g. a value of $1.0$ means a $100\%$ increase in the microbial load. The maximal standard error is 0.74. BH procedure was used at $q=0.1$.
	}
	\label{figure:app_sim_p4_TP_additional_1}
\end{figure}

\begin{figure}[htbp]
	\centering
	\includegraphics[width=1.0\textwidth]{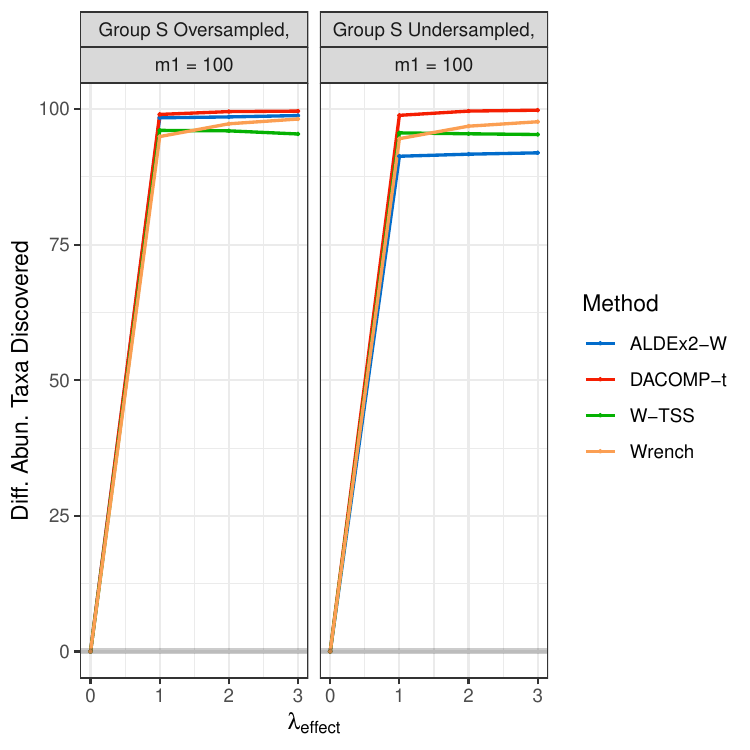}
	\caption{Estimated power of DACOMP and additional competitors for the simulation settings of \S~\ref{app:additional_sim results_varying_seq_depth}. The Y axis represents average number of true discoveries, the X axis represents $\lambda_{effect}$, the increase in percents in the microbial load of a sample with the simulated condition, e.g. a value of $1.0$ means a $100\%$ increase in the microbial load. The maximal standard error is 0.74. BH procedure was used at $q=0.1$.
	}
	\label{figure:app_sim_p4_TP_additional_2}
\end{figure}

Figures \ref{figure:app_sim_p4_TP_additional_1}-\ref{figure:app_sim_p4_TP_additional_2} describe the power of DACOMP and competitors, in a split format similar to Figures~\ref{figure:app_sim_p4_FDR_additional_1}-\ref{figure:app_sim_p4_FDR_additional_2}. The figures detail the statistical power for $m_1=100$. For $m_1=10$, all differentially abundant taxa were discovered by all methods. For $m_1=100,\, \lambda\ge 1 $, we observe DACOMP and DACOMP-t to have the highest statistical power. For the case where "Group Y is Undersampled", the gap in statistical power may be as large as 12 discoveries on average, when comparing DACOMP and ALDEx2-t.

\subsection{A setting where only the rare taxa are differentially abundant}\label{subsec:sim_multinom}

We consider a setting where $supp\left(\bY\right)=\{0,1\}$, and $\vec{\bP}|\{\bY=0\}$ is constant, with the first $m_A$ components having the values $p_{A}/m_{A}$. The remaining $m-m_{A}$ taxa have entries $\left(1-p_{A}\right)/\left(m-m_{A}\right)$:
$$\textbf{P}|\{\textbf{Y}=0\} = \left(\frac{p_{A}}{m_{A}},...,\frac{p_{A}}{m_{A}},\frac{\left(1-p_{A}\right)}{\left(m-m_{A}\right)},...,\frac{\left(1-p_{A}\right)}{\left(m-m_{A}\right)}\right).$$
The parameter $p_{A}$ represents the relative part of the microbial load for the first $m_{A}$ taxa. For values of $0.5\le p_{A}\le 0.9$, the first $m_{A}$ taxa will contain the majority of the microbial load.

Subjects of the second group were multinomial samples with differentially abundant taxa selected from the taxa with relative frequencies $\left(1-p_{A}\right)/\left(m-m_{A}\right)$,
$$\textbf{P}|\{\textbf{Y}=1\} =  \left(1-w\right)\cdot \left(\textbf{P}|\{\textbf{Y}=0\}\right) + w\cdot \left(0,...,0,1,...,1,0,...,0\right),$$
where $w$ is the proportion of signal added to the vector of relative frequencies and vector on the right term has $m_1$ entries with indices larger than $m_{A}$ with a value of $1$, rendering the corresponding taxa as differentially abundant. For each simulated dataset, $40$ samples where sampled, evenly split between $\textbf{Y}=0$ and $\textbf{Y}=1$. The observed count vectors, $\textbf{X}_i$'s, were multinomial random vectors with $N_{reads} = 2500$ reads in each vector and sampled using $\textbf{P}_i | \textbf{Y}_i$ for each observation. We examined simulations with $m=300, m_{A}=30, w=0.35, m_1 \in \{120,60\} , p_{A} \in\{0.9,0.8,0.7,0.6,0.5\}$.

Table \ref{table:SIM_MUTLINOM} shows the estimated FDR, for each simulated setting, by $m_1$ and $p_{A}$. We note that the only methods providing FDR control across all scenarios are DACOMP and HG. Since there is no overdispersion in the data, these two methods are theoretically valid (but DACOMP is also valid when there is overdispersion). ANCOM and ALDEx2-t  provide FDR control for settings with $m_1=60$, but not for settings with $m_1 = 120$. ANCOM's loss of FDR control for settings with $m_1=120$, is related to the loss of power: As described in \S~\ref{subsec:existing_methods}, the method of \citet{mandal2015analysis} makes use of the $p$-values $p_{j,k}$, testing if the ratio between the $j$th and $k$th taxa is associated with the measured trait for every pair of taxa, $j$ and $k$.  Implicitly, it is assumed that if taxon $j$ or $k$ are differentially abundant, the $p$-value of $p_{j,k}$ will be smaller than $\alpha$, e.g., $\alpha  = 0.1$, with high probability. If this assumption is violated, the highest values of $\mathcal{W}_{j}$ may not be obtained by the differentially abundant taxa. The setting generated demonstrates this effect. ANCOM fails to identify the differentially abundant taxa, and instead associates the most abundant taxa with the disease. W-CSS provides FDR control for only two of the scenarios considered. For DACOMP-ratio, the estimated FDR for $m_1=60, p_{A}\in \{0.7,0.8,0.9\}$ is higher than $q=0.1$.

In terms of power, all methods discovered all differentially abundant taxa, expect for: ANCOM discovered 33,30,28,34 and 28 taxa in the settings of rows 1-5, respectively; ALDEx2-t discovered 120,118,116,115 and 111 taxa in the settings of rows 1-5, respectively. The maximum standard error for average number of taxa discovered is 3.71.

\begin{table}[h]
	\caption{Estimated FDR of DACOMP and competitors (Columns 3-8) for the simulations where the most abundant taxa are not differentially abundant. Column 1-2 give the number of differentially abundant taxa and the value of the parameter $p_A$, respectively. The maximum standard error a table entry is 0.03. For DACOMP-ratio, the maximum standard error across table entries is $0.004$.}
	
	\small
	\begin{tabular}{rrllllllll}
		\hline	
		$m_1$ & $p_{A}$ & ALDEx2-t & ANCOM & DACOMP-ratio & DACOMP & HG & W-CSS \\ 
		\hline
		120 & 0.90 & 0.3 & 0.34 & 0.1 & 0.07 & 0.07 & 0.54 \\ 
		120 & 0.80 & 0.46 & 0.25 & 0.09 & 0.07 & 0.07 & 0.57 \\ 
		120 & 0.70 & 0.5 & 0.28 & 0.08 & 0.07 & 0.06 & 0.48 \\ 
		120 & 0.60 & 0.52 & 0.24 & 0.07 & 0.06 & 0.05 & 0.38 \\ 
		120 & 0.50 & 0.54 & 0.24 & 0.07 & 0.06 & 0.06 & 0.34 \\ 
		60 & 0.90 & 0 & 0.01 & 0.14 & 0.09 & 0.1 & 0.59 \\ 
		60 & 0.80 & 0.02 & 0.01 & 0.12 & 0.09 & 0.09 & 0.44 \\ 
		60 & 0.70 & 0.02 & 0.01 & 0.11 & 0.08 & 0.08 & 0.43 \\ 
		60 & 0.60 & 0.01 & 0.01 & 0.1 & 0.08 & 0.07 & 0.1 \\ 
		60 & 0.50 & 0 & 0.01 & 0.09 & 0.07 & 0.07 & 0.08 \\ 
		\hline
	\end{tabular}\label{table:SIM_MUTLINOM}
	
\end{table}

\subsection{Cases with no compositionality}\label{subsec:sim_nocomp}
We wish to asses the potential loss of power by using a method that adjusts for compositionality, when adjustment for compositionality is in fact unnecessary for valid inference. Taxon counts are considered as an independent sample from a negative binomial distribution where the mean is $\mu$ and the variance is given by $\mu + \mu^2/5$. 

Simulated data for samples with trait values of $\textbf{Y}=0$ consisted of $m=1000$ taxa sampled as independent negative binomial variables, with $50$ highly abundant taxa with a mean of $200$, $150$ medium abundance taxa with a mean of $20$ and $800$ taxa with low abundance having a mean of $1$. For simulating samples with a trait values of $\textbf{Y}=1$, 10 taxa with high abundance, 10 taxa with medium abundance, and 30 taxa with low abundance were selected as differentially abundant. Out of each abundance group (means of 1,20,200), of the differentially abundant taxa half had their means reduced by 75\% and half had their means increased by 75\%. Therefore the distribution of non differentially abundant taxa is the same in the two groups. Sample size was $n_0=n_1 \in \{15,20,25,30\}$, with $n_0$ and $n_1$ denoting the number of samples in group. 

In terms of FDR, all methods except HG controlled the FDR at the required rate. This result is expected since all non differentially abundant taxa have maintained their marginal distributions across study groups. For HG, due to overdispersion in the data, the average FDR was $0.68$ or above for all settings considered. Hence, HG was removed from power comparisons.

Table \ref{table:SIM_NOCOMP_TP} describes the estimated power of the different methods. W-CSS discovers the highest number of differentially abundant taxa, followed by DACOMP-ratio. ANCOM and ALDEx2-t have a comparable number of discoveries across all sample sizes considered. DACOMP has higher power compared to both ANCOM and ALDEx2-t, and lower power than DACOMP-ratio. The difference in power between W-CSS and DACOMP-ratio to other competitors results mainly from detecting differentially abundant taxa with low counts.

\begin{table}[htbp]
	\centering
	\caption{Estimated average number of differentially abundant taxa discovered by DACOMP and competitors. The maximum standard error is and $0.42$.}
	\label{table:SIM_NOCOMP_TP}
	\small
	\begin{tabular}{llllll}
		\hline
		$n_X$:$n_Y$ & ALDEx2-t & ANCOM &DACOMP-ratio & DACOMP & W-CSS \\ 
		\hline
		15:15 & 10.39 & 11 & 15.44 & 12.2 & 17.34 \\ 
		20:20 & 13.58 & 12.42 & 19.62 & 14.63 & 22.18 \\ 
		25:25 & 15.48 & 13.91 & 23.38 & 16.78 & 26.91 \\ 
		30:30 & 16.91 & 16.43 & 28.27 & 19.4 & 31.14 \\ 
		\hline
	\end{tabular}
\end{table}

\subsection{Simulation results for additional methods}\label{app:additional competitors}
In this appendix we present results for the simulation study discussed in \S~\ref{sec:simulations}, for the following methods: W-TSS, ALDEx2-W, DACOMP-t and WRENCH. Method description and details are at the start of \S~\ref{sec:simulations}. In terms of Power and FDR, unless stated otherwise, W-TSS was similar to  W-CSS, ALDEx2-W was similar to ALDEx2-t and DACOMP-t was similar to DACOMP. Subsection \S~\ref{app:additional_sim results_varying_seq_depth} focuses on settings where the sequencing depth of samples differed across study groups.

\begin{figure}[htbp]
	
	\centering
	\includegraphics[width=1.0\textwidth]{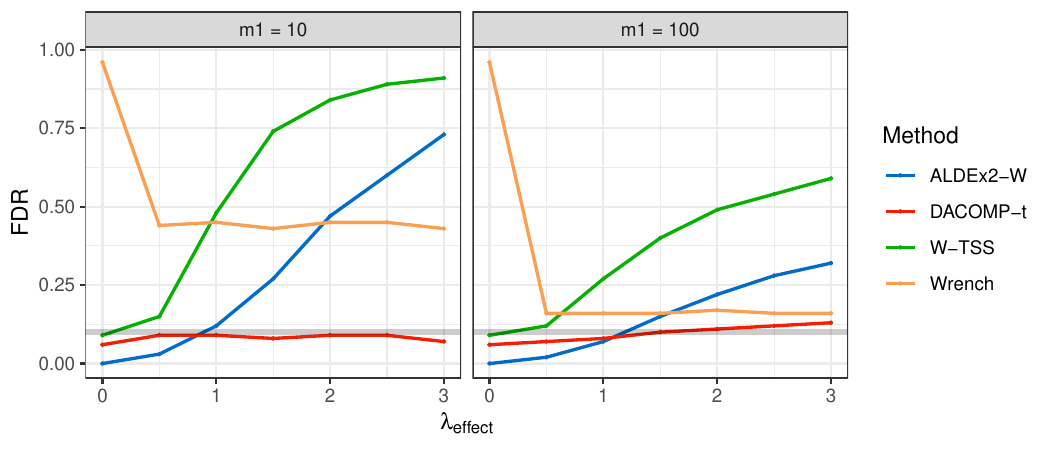}
	\caption{Estimated FDR of DACOMP and competitors for the simulation settings of \S~\ref{subsec:sim_gut}. The Y axis represents estimated FDR, the X axis represents $\lambda_{effect}$, the increase in percents in the microbial load of a sample with the simulated condition, e.g. a value of $1.0$ means a $100\%$ increase in the microbial load. The maximal standard error is 0.04. BH procedure was used at $q=0.1$. The gray line marks the value $q=0.1$.
	}
	\label{figure:app_sim_p1_FDR_additional}
\end{figure}

Figure~\ref{figure:app_sim_p1_FDR_additional} shows the estimated FDR for the additional methods listed above, obtained the simulation settings discussed in \S~\ref{subsec:sim_gut}. Similar to DACOMP, DACOMP-t is shown to control the false discovery rate at $q=0.1$. W-TSS does not control the false discovery rate for all scenarios with $\lambda_{effect}>=1.0$, since it provides marginal inference alone. ALDEx2-W failed to control for false positives at $\lambda_{effect}>=1.5$. When comparing ALDEx2-W to ALDEx2-t in terms of FDR control, FDR rates for ALDEx2-W were significantly higher, e.g., for $\lambda_{effect}=2$ with $m_1=100$, the estimated FDR for ALDEx2-W was $0.22$ but was only $0.12$ for ALDEx2-t. For WRENCH, FDR was not controlled under the global null, $\lambda_{effect}=0$, or for $m_1$ = 10. For $m_1=100,\lambda_{effect}>0$, the standard error for WRENCH's FDR estimates were smaller than 0.01, indicating that the observed FDR levels, approximately 0.16 across all values of $\lambda_{effect}>0$, are significantly different from $q=0.1$. The lack of FDR control could be related to the warning message about failure of algorithm convergence. The method was run with the default parameters. Other parameter estimates for WRENCH may produce better FDR control.

\begin{figure}[htbp]
	
	\centering
	\includegraphics[width=1.0\textwidth]{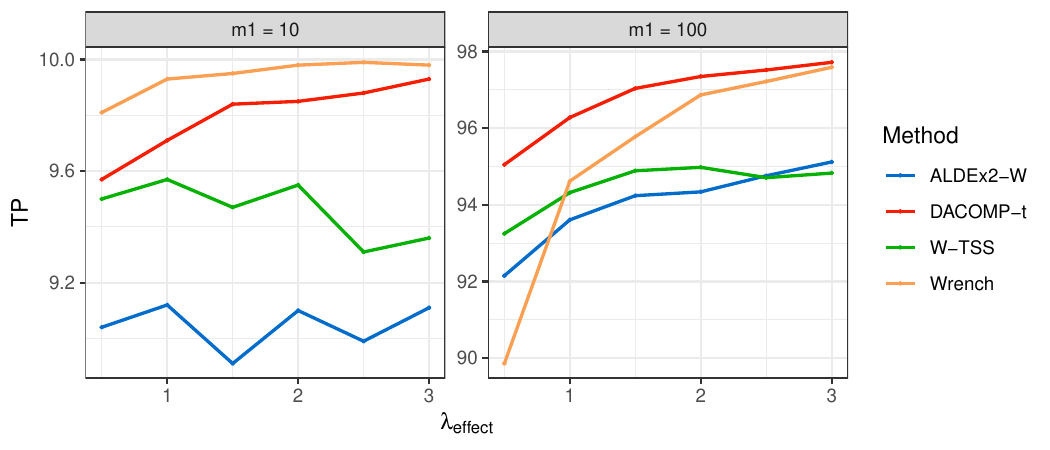}
	\caption{Estimated power of DACOMP and competitors for the simulation settings of \S~\ref{subsec:sim_gut}. The Y axis represents average number of true discoveries, the X axis represents $\lambda_{effect}$, the increase in percents in the microbial load of a sample with the simulated condition, e.g. a value of $1.0$ means a $100\%$ increase in the microbial load. The maximal standard error is 0.73. BH procedure was used at $q=0.1$.
	}
	\label{figure:app_sim_p1_TP_additional}
\end{figure}

Figure~\ref{figure:app_sim_p1_TP_additional} shows the power of DACOMP-t and alternative methods, for the scenarios of \S~\ref{subsec:sim_gut}. All method variants (DACOMP-t,ALDEx2-W, W-TSS) gave results similar to their respective variants, as described in the beginning of the section.

Table~\ref{table:SIM_MUTLINOM_APPENDIX} shows the estimated FDR for DACOMP-t and competitors, for the simulation scenarios discussed in \S~\ref{subsec:sim_multinom}. DACOMP-t is the only method shown to provide FDR control across all scenarios, similar to DACOMP in Table~\ref{table:SIM_MUTLINOM}. In terms of power, for the scenario in rows 1-5, all methods discovered all differentially abundant taxa, except for ALDEx2-W which discovered 120,117,115,113,109 of the differentially abundant taxa; . The maximum standard error for average number of taxa discovered is 3.71.

\begin{table}[htbp]
	\caption{Estimated FDR of DACOMP-t and competitors (Columns 3-6) for the simulations where the most abundant taxa are not differentially abundant. Column 1-2 give the number of differentially abundant taxa and the value of the parameter $p_A$, respectively. BH procedure was used at $q=0.1$. The maximum standard error a table entry is 0.03.}
	\label{table:SIM_MUTLINOM_APPENDIX}
	\centering
	\begin{tabular}{rrllll}
		\hline
		$m_1$ & $p_{A}$ & ALDEx2-W & DACOMP-t & W-TSS & WRENCH \\ 
		\hline
		120 & 0.90 & 0.42 & 0.07 & 0.34 & 0.46 \\ 
		120 & 0.80 & 0.55 & 0.07 & 0.43 & 0.56 \\ 
		120 & 0.70 & 0.57 & 0.07 & 0.49 & 0.57 \\ 
		120 & 0.60 & 0.57 & 0.06 & 0.53 & 0.57 \\ 
		120 & 0.50 & 0.57 & 0.06 & 0.55 & 0.58 \\ 
		60 & 0.90 & 0 & 0.09 & 0.52 & 0.46 \\ 
		60 & 0.80 & 0.01 & 0.09 & 0.64 & 0.65 \\ 
		60 & 0.70 & 0.01 & 0.08 & 0.71 & 0.71 \\ 
		60 & 0.60 & 0.01 & 0.08 & 0.74 & 0.73 \\ 
		60 & 0.50 & 0 & 0.08 & 0.76 & 0.75 \\ 
		\hline
	\end{tabular}
\end{table}

Table~\ref{table:SIM_NOCOMP_FDR_APPENDIX} presents the estimated FDR for DACOMP-t and alternative methods, for the scenarios described in \S~\ref{subsec:sim_nocomp}. For the scenarios of \S~\ref{subsec:sim_nocomp}, non-differentially abundant taxa maintained their marginal distributions of counts across study groups. Hence, all tests presented in Table~\ref{table:SIM_NOCOMP_FDR_APPENDIX} provide valid FDR control.

\begin{table}[htbp]
	\caption{Estimated FDR of DACOMP-t and competitors (Columns 2-5) for simulations with no compositionality, for various sample sizes (Column 1). BH procedure was applied at level  $q=0.1$. The maximum standard error of a table entry is $0.01$.}
	\label{table:SIM_NOCOMP_FDR_APPENDIX}
	\centering
	\begin{tabular}{lllll}
		\hline
		$n_X$:$n_Y$ & ALDEx2-W & DACOMP-t & W-TSS & WRENCH \\ 
		\hline
		15:15 & 0 & 0.05 & 0.09 & 0.06 \\ 
		20:20 & 0 & 0.08 & 0.09 & 0.06 \\ 
		25:25 & 0 & 0.09 & 0.09 & 0.06 \\ 
		30:30 & 0 & 0.07 & 0.1 & 0.07 \\ 
		\hline
	\end{tabular}
\end{table}

Table~\ref{table:SIM_NOCOMP_TP_APPENDIX} shows the number of true positive discoveries for DACOMP-t and alternative methods, for the scenarios of \S~\ref{subsec:sim_nocomp}. WRENCH is shown to provide the highest power, even higher than W-TSS. 

\begin{table}[htbp]
	
	\caption{Average number of differentially abundant taxa discovered by DACOMP-t and competitors that controlled FDR (Columns 2-5) for simulations with no compositionality, by sample size (Column 1). BH procedure was applied at level $q=0.1$. The maximum standard error of a table entry is $0.42$.}
	\label{table:SIM_NOCOMP_TP_APPENDIX}
	\centering
	\begin{tabular}{lllll}
		\hline
		$n_X$:$n_Y$ & ALDEx2-t & DACOMP-t & W-TSS & WRENCH \\ 
		\hline
		15:15 & 11.48 & 12.35 & 16.97 & 18.22 \\ 
		20:20 & 13.52 & 14.81 & 22.5 & 22.82 \\ 
		25:25 & 15.6 & 16.73 & 27.08 & 27.78 \\ 
		30:30 & 17.18 & 19.84 & 30.86 & 31.78 \\ 
		\hline
	\end{tabular}
\end{table}

\section{Additional results for the Crohn's disease data example}\label{app:additional_CD_results}

\begin{figure}[htbp]
	\begin{subfigure}[t]{1.0\textwidth}
		\centering
		\includegraphics[width=0.6\linewidth]{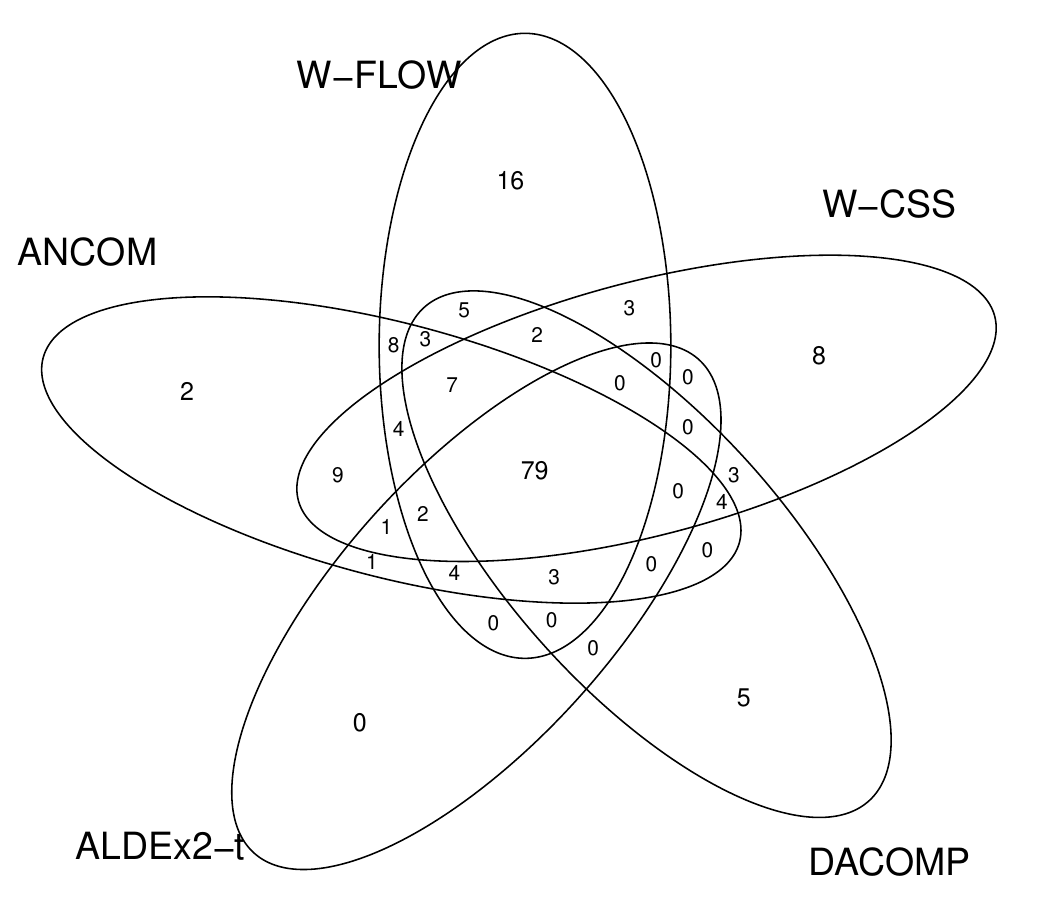}
		\caption{}
	\end{subfigure}
	
	\caption{Graphical representation of discoveries shared by different methods for the 264 taxa, with at least 10 counts, on average, per sample. }
	\label{fig:Crohn_shared_Venn_Abundant}
\end{figure}

\begin{figure}[htbp]
	\begin{subfigure}[t]{1.0\textwidth}
		\centering
		\includegraphics[width=0.6\linewidth]{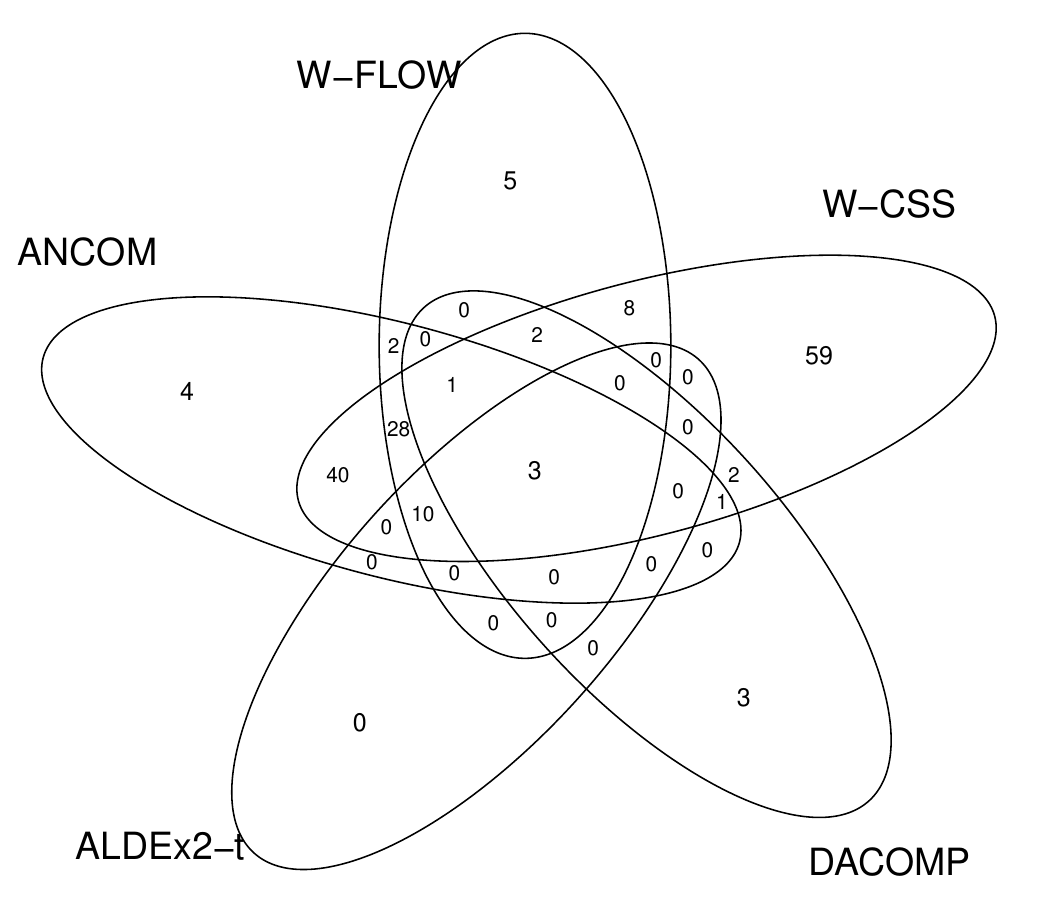}
		\caption{}
	\end{subfigure}
	
	\caption{Graphical representation of discoveries shared by different methods for the 1305 'rare' taxa, with less than 10 counts, on average, per sample.}
	\label{fig:Crohn_shared_Venn_Rare}
\end{figure}

Figures \ref{fig:Crohn_shared_Venn_Abundant} and \ref{fig:Crohn_shared_Venn_Rare} present a Venn diagram of the shared discoveries, for the methods compared in the Crohn's disease data example of \S~\ref{sec:RDE_GUT}, for the abundant and rare taxa, respectively. Taxa with at least 10 counts on average, per sample, were considered abundant. Out of 1569 taxa, 264 were considered abundant. See \S~\ref{sec:RDE_GUT} for a detailed discussion of the results. 

Table \ref{table:RDE_GUT_parameter} shows the number of discoveries by DACOMP for several values of $S_{crit}$ alongside the obtained reference size and the number of discoveries shared with other methods. For the values of $S_{crit}$ described in the table, as $S_{crit}$ increases, more taxa enter the selected set of references. As a result, less taxa are tested and discovered as differentially abundant. While the number of taxa discovered as differentially depends on $S_{crit}$, the difference in the number of discoveries between rows of Table \ref{table:RDE_GUT_parameter} is minor compared to the 70-150 additional taxa that may be considered differentially abundant by using ANCOM, W-CSS or W-FLOW.

\begin{table}[ht]
	\caption{Number of discoveries by $S_{crit}$ for DACOMP. Columns 2-5 show for each value of $S_{crit}$ the number of discoveries shared with ANCOM, W-FLOW, and ALDEx2-t, and the number of OTUs in the selected reference set $B$, respectively.}
	\label{table:RDE_GUT_parameter}
	\small
	\begin{tabular}{rrrrrrr}
		\hline
		$S_{crit}$ & Discoveries & Shared, ANCOM & Shared, W-FLOW &Shared, ALDEx2-t & $|B|$ \\ 
		\hline
		1.2 & 149 &  121 &  122&92 & 1221   \\ 
		1.3 & 123 &  101 &  105&85 & 1288   \\ 
		1.4 & 108 &  93  &  98&79  & 1335   \\ 
		\hline
	\end{tabular}
\end{table}

\section{Further examination of the reference selection procedure}\label{app:appendix_B}
In this appendix we further examine the reference selection procedure suggested in \S~\ref{subsec:choosing_reference} and alternative reference selection procedures. In \S~\ref{app:ref_score_distribution} we detail how the tuning parameter of $S_{crit}$ was selected and examine the chance of a differentially abundant taxon to erroneously be inserted into the selected reference set. In \S~\ref{app:naive reference selection} we examine the FDR of naive reference selection methods, e.g., picking the reference set of taxa at random. In \S~\ref{app:reference_validity} we propose a procedure for checking the validity of a reference set of tax.

\subsection{Selecting $S_{crit}$}\label{app:ref_score_distribution}

The data adaptive method for reference selection presented in \S~\ref{subsec:choosing_reference} has a single tuning parameter, $S_{crit}$. Taxa with a reference score below the parameter $S_{crit}$ constitute the reference set. The value of $S_{crit}$ was set to $1.3$ in \S~\ref{sec:simulations} - \S~\ref{sec:RDE_GUT} after observing the distribution of reference scores in real and simulated data. 

\begin{figure}[htbp]
	
	\begin{subfigure}[t]{.45\textwidth}
		\centering
		\includegraphics[width=\linewidth]{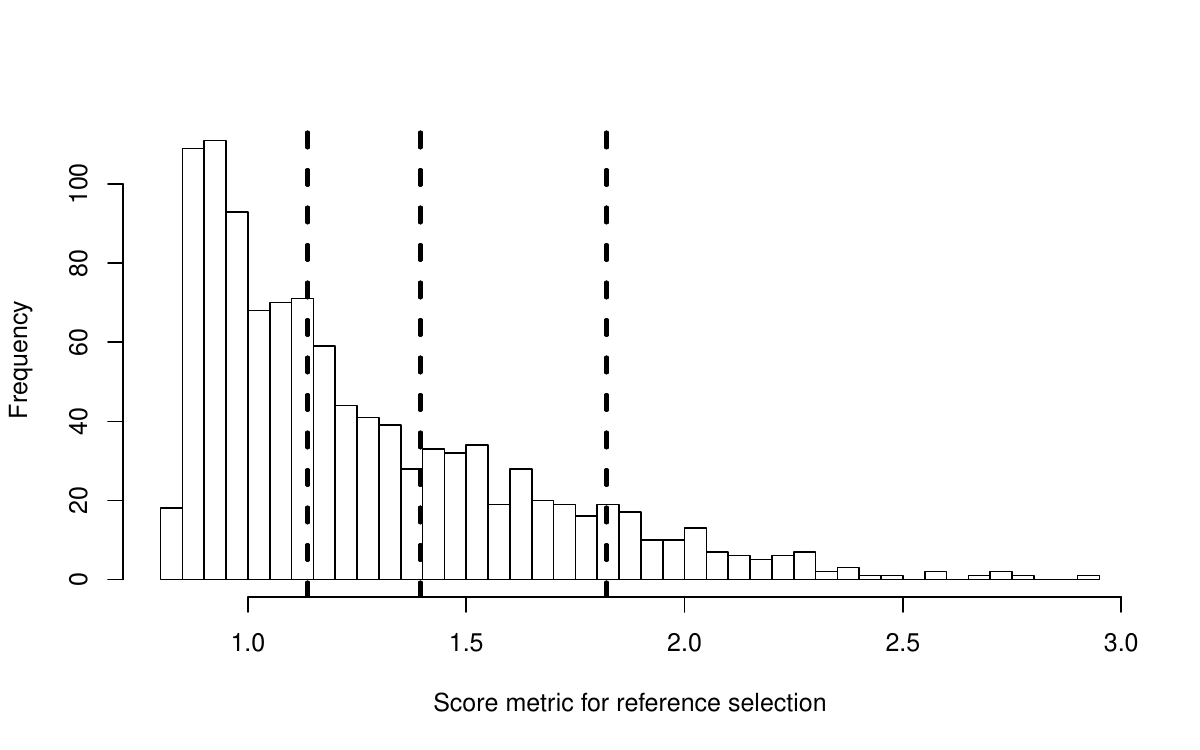}
		\caption{Case $m_1=0$ from the simulation of \S~\ref{subsec:sim_gut}}
	\end{subfigure}
	\hfill
	\begin{subfigure}[t]{.45\textwidth}
		\centering
		\includegraphics[width=\linewidth]{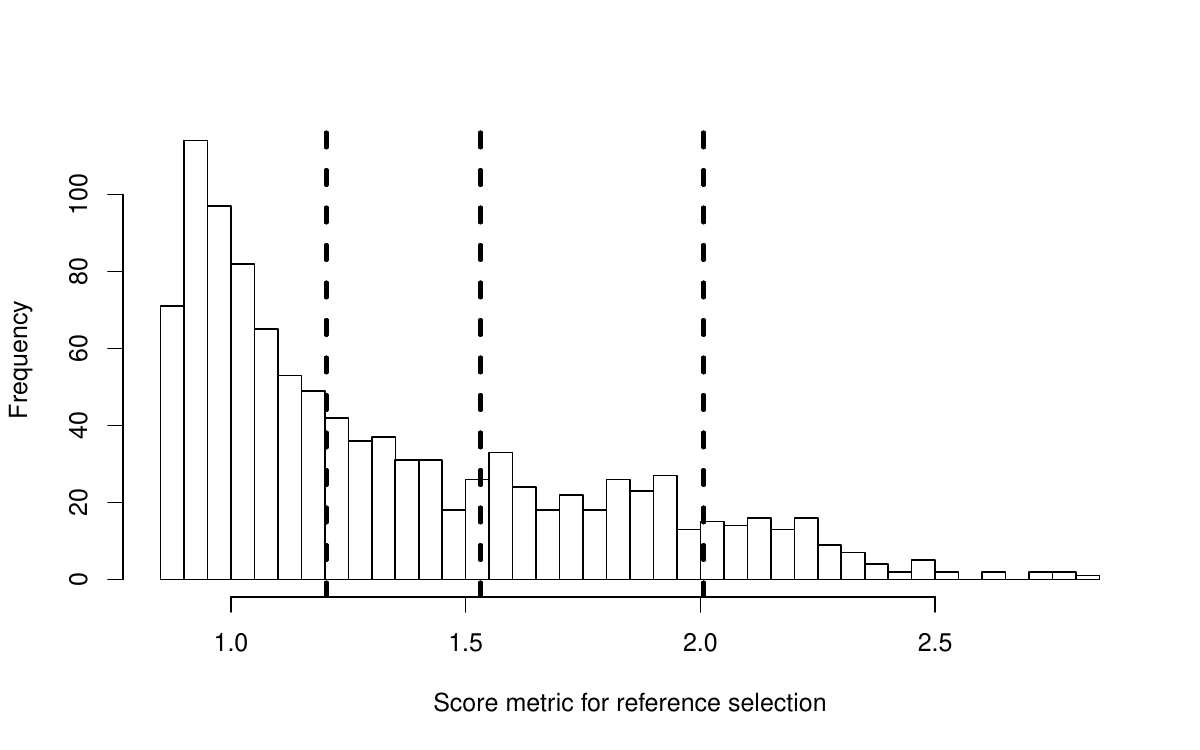}
		\caption{Case $m_1=100,\lambda_{effect}=0.5$ from the simulation of \S~\ref{subsec:sim_gut} }
	\end{subfigure}
	
	\medskip
	
	\begin{subfigure}[t]{.45\textwidth}
		\centering
		\includegraphics[width=\linewidth]{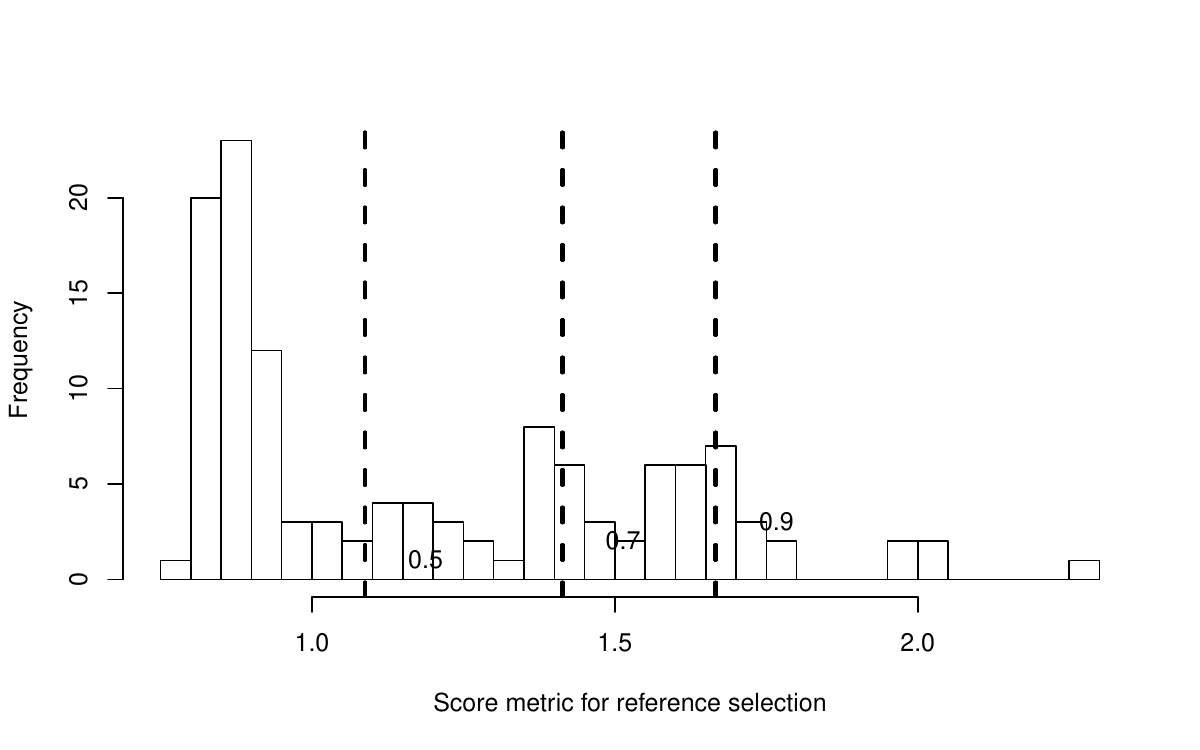}
		\caption{Comparison of Hard Palate and Subgingival Plaque in \S~\ref{app:RDE_HMP} }
	\end{subfigure}
	\hfill
	\begin{subfigure}[t]{.45\textwidth}
		\centering
		\includegraphics[width=\linewidth]{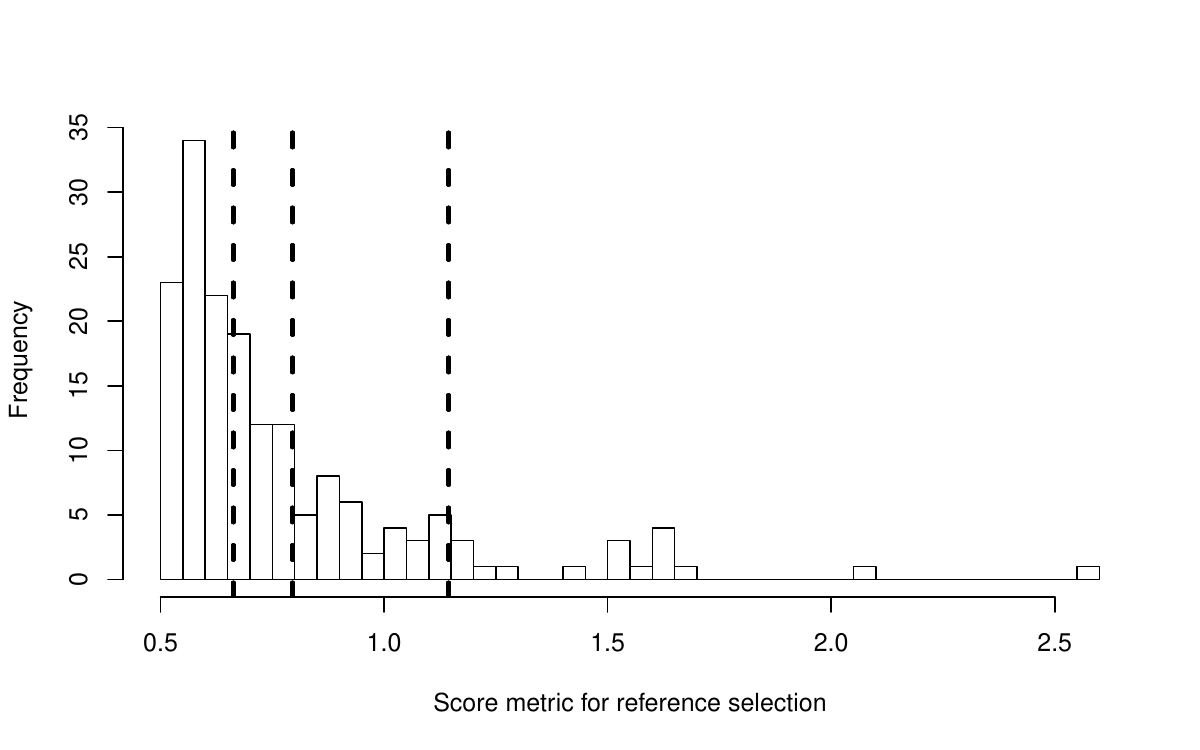}
		\caption{Comparison of left and right Retroauricular Crease in \S~\ref{app:RDE_HMP} }
	\end{subfigure}
	\caption{Histograms for reference scores computed in selected simulations and data analyses. Median and 0.7, 0.9 percentiles for reference scores are presented using vertical dashed lines in each subplot.}
	\label{app:ref_score_fig}
\end{figure}

Figure~\ref{app:ref_score_fig} shows the distribution of reference scores for several real and simulated data sets. Values of $S_{crit}$ in the range $\left[1.0,1.4\right]$ select roughly 60-70\% of taxa as a reference set. The remaining portion of taxa exhibit reference scores which are substantially higher than $1.3$, and are not valid candidates to form the reference set $B$. Subplot (d) shows a relatively large portion of taxa with reference scores below $1.3$. However, the comparison in subplot (d) is between the left and right Retroauricular creases. If any taxa are differentially abundant between the two sites, it is plausible to believe their number is small. Hence, the extreme values of the distribution in subplot $D$ hint at $S_{crit}=1.3$ as a plausible threshold as well. 

Table~\ref{app:table_bad_ref} shows the mean number of differentially abundant taxa inserted into the reference set. This includes the scenarios of \S~\ref{subsec:sim_multinom}-\S~\ref{subsec:sim_nocomp}, where the signal present in differentially abundant taxa is much smaller than \S~\ref{subsec:sim_gut}. For most scenarios, no differentially abundant taxa have entered the reference set. We see that for some scenarios, some of the differentially abundant taxa have entered the reference set, however this occurred in a small fraction of the cases, with the mean number of differentially abundant included in the reference set being less than 1. Moreover, control of the FDR was not compromised in those settings.

\begin{table}[htbp]
	\centering
	\caption{Mean number of differentially abundant (DA) taxa inserted into the reference set, by simulation scenario. The standard error across 100 simulated data sets is given in brackets.}
	\label{app:table_bad_ref}
	\begin{tabular}{lcc}
		\hline
		Simulation case & Mean number of DA taxa in $B$\\ 
		\hline
		\S~\ref{subsec:sim_gut}, $m_1=0$ & 0 (0) \\ 
		\S~\ref{subsec:sim_gut}, $m_1=10, \lambda_{effect} = 0.5$ & 0 (0) \\ 
		\S~\ref{subsec:sim_gut}, $m_1=100, \lambda_{effect} = 0.5$ & 0.49 (0.05) \\ 
		\S~\ref{subsec:sim_gut}, $m_1=10, \lambda_{effect} = 1.0$ & 0 (0) \\ 
		\S~\ref{subsec:sim_gut}, $m_1=100, \lambda_{effect} = 1.0$ & 0.67 (0.05) \\ 
		\S~\ref{subsec:sim_gut}, $m_1=10, \lambda_{effect} = 1.5$ & 0 (0) \\ 
		\S~\ref{subsec:sim_gut}, $m_1=100, \lambda_{effect} = 1.5$ & 0.74 (0.06) \\ 
		\S~\ref{subsec:sim_gut}, $m_1=10, \lambda_{effect} = 2.0$  & 0 (0) \\ 
		\S~\ref{subsec:sim_gut}, $m_1=100, \lambda_{effect} = 2.0$  & 0.8 (0.06) \\ 
		\S~\ref{subsec:sim_gut}, $m_1=10, \lambda_{effect} = 2.5$ & 0 (0) \\ 
		\S~\ref{subsec:sim_gut}, $m_1=100, \lambda_{effect} = 2.5$ & 0.9 (0.07) \\ 
		\S~\ref{subsec:sim_gut}, $m_1=10, \lambda_{effect} = 3.0$ &  0 (0) \\ 
		\S~\ref{subsec:sim_gut}, $m_1=100, \lambda_{effect} = 3.0$ &  0.98 (0.06) \\ 
		
		\S~\ref{subsec:sim_multinom}, Case 1 & 0 (0) \\ 
		\S~\ref{subsec:sim_multinom}, Case 2 & 0 (0) \\ 
		\S~\ref{subsec:sim_multinom}, Case 3 & 0 (0) \\ 
		\S~\ref{subsec:sim_multinom}, Case 4 & 0 (0) \\ 
		\S~\ref{subsec:sim_multinom}, Case 5 & 0 (0) \\ 
		\S~\ref{subsec:sim_multinom}, Case 6 & 0 (0) \\ 
		\S~\ref{subsec:sim_multinom}, Case 7 & 0 (0) \\ 
		\S~\ref{subsec:sim_multinom}, Case 8 & 0 (0) \\ 
		\S~\ref{subsec:sim_multinom}, Case 9 & 0 (0) \\ 
		\S~\ref{subsec:sim_multinom}, Case 10 & 0 (0) \\ 
		\S~\ref{subsec:sim_nocomp}, Case 1 & 0.41 (0.07) \\ 
		\S~\ref{subsec:sim_nocomp}, Case 2 & 0.46 (0.08) \\ 
		\S~\ref{subsec:sim_nocomp}, Case 3 & 0.39 (0.07) \\ 
		\S~\ref{subsec:sim_nocomp}, Case 4 & 0.46 (0.08) \\ 
		\hline
	\end{tabular}
\end{table}

\subsection{Examining naive approaches for reference selection}\label{app:naive reference selection}
The reference selection method presented in \S~\ref{subsec:choosing_reference} aims to find a set of non-differentially abundant taxa. In this subsection, we show how selecting references at random, or while disregarding \eqref{eq:NullAssumption}, can lead to lack of FDR control by the method presented in \S~\ref{subsec:oracle_test}.

We examine two possible alternative approaches for the reference selection method \S~\ref{subsec:choosing_reference}. The first approach picks taxa at random for the reference set. The second approach picks the most abundant taxa as a reference set. Taxon abundance is computed by the total number of counts observed in a taxon across all subjects. In order to evaluate these approaches, we performed the following evaluation: for a given simulation setting, e.g., the 5th setting presented in \S~\ref{subsec:sim_multinom}, 200 data sets were sampled. For each realized dataset, two reference set of taxa were selected using the approaches stated above. The method proposed in \S~\ref{subsec:oracle_test} was used to detect differentially abundant taxa using the selected reference sets. The BH procedure was applied for FDR control at level $q=0.1$. 

Table~\ref{app:naive_ref_results} presents the estimated FDR of the two alternative reference selection methods by scenario. Both procedures are observed to select a large number of differentially abundant taxa into the reference set $B$. As a result, the procedure of \S~\ref{subsec:oracle_test} lacks FDR control. 

\begin{table}[htbp]
	\centering
	\caption{Estimated FDR for naive reference selection methods, across selected scenarios. RAND stands for picking 50 taxa at random as $B$. ABUND stands for picking the 50 most abundant taxa as differentially abundant. Entries significantly higher than 0.1 are marked with a *.}
	\label{app:naive_ref_results}
	\begin{tabular}{lll}
		\hline
		Scenario & RAND & ABUND \\ 
		\hline
		\S~\ref{subsec:sim_gut}, $m_1=100,\lambda_{effect}=0.5$ & 0.19* & 0.06  \\ 
		\S~\ref{subsec:sim_gut}, $m_1=10,\lambda_{effect}=2.5$ & 0.34* & 1.00* \\ 
		\S~\ref{subsec:sim_multinom}, Case 5 & 0.53* & 0.17* \\  
		\S~\ref{subsec:sim_nocomp}, Case 4 & 0.19* & 0.09 \\  
		\hline
	\end{tabular}
\end{table}

\subsection{Checking the validity of a reference set of taxa}\label{app:reference_validity}

For a valid set of reference taxa, denoted by the entries with 'ones' in the binary vector $\b b$,the relation given by \eqref{eq:NullAssumption} should hold:
\begin{equation}\label{eq:ref_test}
H_0^{\textbf{b}}: \frac{\left(\textbf{P}(\textbf{b})\right)}{\textbf{b}'\textbf{P}}\indep\textbf{Y}.
\end{equation}
Tests for \eqref{eq:ref_test} test the validity of the reference set of taxa: if the reference set of taxa defined by the non-zero entries of $\textbf{b}$ is comprised solely of non differentially abundant taxa, then  \eqref{eq:ref_test} holds. A simple test for \eqref{eq:ref_test} is the following: (1) From each sample, select the sub-vector of reference taxa given by the indices $\left(b_1,b_2,...,b_r\right)$ (2) Rarefy all sub-vectors of reference taxa across samples to uniform depth (3) Test for equality of distributions over the rarefied sub-vectors, using a multivariate test for equality of distributions, e.g., the tests of \citet{anderson2001new} or \citet{heller2012consistent}. This procedure is assumption-free, and only requires selection of a distance metric for computing pairwise distances between samples. We will denote this procedure as a reference validation procedure, or RVP.

In order to examine the validity of this procedure, we conduct a simulation study. If the proposed RVP is valid, and no differentially abundant taxa have entered the reference set, the probability of the RVP to reject its null hypothesis should match the nominal Type I error rate used for testing. A higher probability to reject the RVP's null hypothesis indicates a problem in the proposed procedure. For a given simulation setting, e.g. case 1 from \S~\ref{subsec:sim_multinom}, we sample $1000$ datasets. For each sampled dataset, we select references according to the method presented in \S~\ref{subsec:choosing_reference} with $S_{crit}=1.3$. We carry out the reference validation procedure suggested above, for all data realizations in which the reference set of taxa contains no differentially abundant taxa. As a multivariate test for equality of distributions, we use several options for each sampled data set: the HHG test of \citet{heller2012consistent}, the PERMANOVA test of \citet{anderson2001new}, and the DISCO test of \citet{rizzo2010disco}. As a distance metric to be used by the suggested tests, we use the L2 and L1 distances and the Bray-Curtis dissimilarity metric. Overall, 9 variations of the above procedure are considered. Multivariate tests are performed at level $\alpha=0.1$. For this simulation study, we considered only the settings whose effect size was either the smallest or the largest in the respective subsection, specifically: simulation cases from \S~\ref{subsec:sim_gut} with $\lambda_{effect}\in\{0.5,3.0\}$, and simulation cases 1, 5, 6, 10 from \S~\ref{subsec:sim_multinom}. The simulation settings of \S~\ref{subsec:sim_nocomp} have a non-zero chance for selecting a reference set with a single taxon. For a reference set of taxa comprised of a single taxon, the RVP cannot be carried out. Hence, the settings of \S~\ref{subsec:sim_nocomp} are excluded from this simulation study.

Table~\ref{app:table_GN_T1E} describes the probability estimates of the RVP test to reject the null hypothesis, based on different multivariate tests, distance metrics and simulations cases. Most table entries are within 2 standard errors of the nominal error rate, with the exception of the probability estimates obtained for the HHG test in cases 5 and 10 of \S~\ref{subsec:sim_multinom}, and the HHG based test with the L2 distance metric for the simulation setting with $m_1 = 100,\lambda_{effect}=3.0$ in \S~\ref{subsec:sim_gut}. 
The inflated false-positive rate in some of the scenarios indicates that while the reference set of taxa was selected without considering the group labeling of observations, the counts vectors are not independent of the group labeling. This dependence is discovered when using a multivariate test of independence with a distance metric between count vectors. This inflation in T1E could be avoided if the data used in the RVP is independent of the data used for selecting the reference set of taxa. However, while counts vectors for reference taxa not exactly independent of the group labeling, we found empirically that the procedure of \S~\ref{subsec:oracle_test} provides adequate FDR control in these settings. 

\begin{table}[htbp]
	\centering
	\caption{Probability to reject the null hypothesis in the RVP procedure proposed in \S~\ref{app:reference_validity}. Column 1 describes the simulation setting. Columns 2-7 describe the chance to reject the null hypothesis according to multivariate test used (HHG, DISCO, PERMANOVA) and distance metric (L2 and L1 distances). The maximal standard error for a table entry is 0.02. Testing is done at level $\alpha = 0.1$. Probability estimates significantly different from $0.1$ are marked in grey.}
	
	\label{app:table_GN_T1E}
	\small
	\begin{tabular}{l|c|c|c|c|c|c|}
		\hline
		Scenario & \multicolumn{2}{c}{HHG} & \multicolumn{2}{c}{ENERGY} & \multicolumn{2}{c}{PERMANOVA} \\
		\hline
		& L2 & L1 & L2 & L1 & L2 & L1 \\
		\S~\ref{subsec:sim_gut}, $m_1 = 10,\lambda_{effect}=0.5$ & 0.09 & 0.10 & 0.11 & 0.11 & 0.11 & 0.11 \\		
		\S~\ref{subsec:sim_gut}, $m_1 = 100,\lambda_{effect}=0.5$ & 0.10 & 0.11 & 0.11 & 0.11 & 0.12 & 0.13 \\		
		\S~\ref{subsec:sim_gut}, $m_1 = 10,\lambda_{effect}=3.0$ & 0.10 & 0.10  & 0.10 & 0.09 & 0.10 & 0.09 \\		
		\S~\ref{subsec:sim_gut}, $m_1 = 100,\lambda_{effect}=3.0$ & \cellcolor{lightgray} 0.16 & 0.10 & 0.13 & 0.10 & 0.10 & 0.11 \\	
		\S~\ref{subsec:sim_multinom}, Case 1 & 0.08 & 0.09 & 0.04 & 0.04 & 0.03 & 0.03 \\		
		\S~\ref{subsec:sim_multinom}, Case 5 & \cellcolor{lightgray}0.20 & \cellcolor{lightgray} 0.19 & 0.12 & 0.11 & 0.11 & 0.12 \\		
		\S~\ref{subsec:sim_multinom}, Case 6 & 0.08 & 0.09 & 0.04 & 0.04 &  0.04 & 0.03 \\		
		\S~\ref{subsec:sim_multinom}, Case 10 & \cellcolor{lightgray}0.20 & \cellcolor{lightgray}0.20 & 0.11 & 0.10 & 0.11 & 0.11\\		
	\end{tabular}
	
\end{table}

\section{Simulations for control of type I error under the global null}\label{app:ANCOM_global_null}
In order to estimate the control over false discoveries in ANCOM under the global null, i.e. no differentially abundant taxa, we simulated datasets with taxon counts independently sampled from $pois\left(\mu\right)$ across $m$ taxa. We considered two equal groups ,$n_X,n_Y \in \{50,100\}$, $m\in \{50,100\}$, and $\mu \in \{30,60\}$.

ANCOM has several parameters used in its empirical decision rule. One of the parameters, \verb|multcorr| specifies the type of multiple comparison correction used. ANCOM is highly sensitive to changes in this parameter. \verb|multcorr| may receive one of three values, as follows:

\begin{itemize}
	\item \verb|multcorr = 3| : The matrix of $P$-values used, $P_{j,k}$ as defined in \S~\ref{subsec:existing_methods}, is not corrected for multiplicity. This is the default software parameter.
	\item \verb|multcorr = 2| : The values of $P_{j,k}$ are substituted row-by-row, by their adjusted $P$-values given by the BH procedure. 
	\item \verb|multcorr = 1| : The values of $P_{j,k}$ are substituted  by their BH adjusted $P$-values, with correction for multiplicity done over all ${m \choose 2}$ $P$-values.
\end{itemize}

Testing and multiplicity correction was done at $\alpha = q = 0.05$. All other ANCOM parameters were set to default values. Table~\ref{app:ANCOM_FWER_TABLE} gives the estimate of erroneously rejecting the global null hypothesis for ANCOM across the different settings and values of \verb|multcorr|. The main result is that ANCOM fails to control the false positive rate across all scenarios under the global null, with parameters $\verb|multcorr| = 2$ and $\verb|multcorr| = 3$.

\begin{table}[htbp]
	\centering
	\caption{Probability estimates of ANCOM to erroneously declare taxa as differentially abundant. Counts data generated as independent $pois\left(\mu\right)$, for $m$ taxa, and equal sample sizes $n_X=n_Y$. Columns 4-6 give T1E estimates  by value for parameter 'multcorr'.  T1E level was set in software to $\alpha = 0.05$ Estimates are across 200 repetitions, maximum standard error is  $0.035$.}
	\label{app:ANCOM_FWER_TABLE}
	\begin{tabular}{rrrrrr}
		\hline
		$\mu$ & $m$ & $n_X, n_Y$ & \verb|multcorr = 1| & \verb|multcorr = 2| & \verb|multcorr = 3| \\ 
		\hline
		30 & 50 & 50 & 0.00 & 0.36 & 1.00 \\ 
		60 & 50 & 50 & 0.00 & 0.36 & 0.99 \\ 
		30 & 100 & 50 & 0.00 & 0.51 & 1.00 \\ 
		60 & 100 & 50 & 0.00 & 0.54 & 1.00 \\ 
		30 & 50 & 100 & 0.00 & 0.38 & 1.00 \\ 
		60 & 50 & 100 & 0.00 & 0.30 & 0.99 \\ 
		30 & 100 & 100 & 0.00 & 0.48 & 1.00 \\ 
		60 & 100 & 100 & 0.00 & 0.49 & 1.00 \\ 
		\hline
	\end{tabular}
\end{table}

\section{Comparing adjacent body sites in the Human Microbiome Project}\label{app:RDE_HMP}

The Human Microbiome Project \citep{gevers2012human} is a joint collaboration aimed at studying the behavior of microbial ecologies across the human body. 16S profiles of 300 subjects were sampled at 15-18 body sites, with sampling locations being in the oral cavity, skin sites across the body, airways, vagina and fecal samples. We wish to analyze the differences in microbiome composition at adjacent body sites.
The OTU table and taxonomy available from by the link given in \citet{kumar2018analysis} contains 4788 samples and 45383 OTUs. Since OTU picking was done for all body sites combined, many OTUs are prevalent at a small portion of body sites. See \citet{kumar2018analysis} for a comprehensive comparison of normalization approaches with this dataset.

OTUs in the data are associated with a taxon in the known common taxonomy of Kingdom-Phylum-Class-Order-Family-Genus-Species. Some OTUs are associated with a known species of bacteria while others are associated with a high level taxon such as a Genera or Family. Moreover, several OTUs may be linked to the same taxonomic affiliation as a single species may have several known 16S sequences.

To reduce the dimensionality of the data, OTUs counts were aggregated to the Genus level. All OTUs with the same Genus affinity were aggregated to the same vector index. OTUs whose taxonomic affiliation was higher than Genus, were aggregated by their closest affinity, i.e. all OTUs which had Family identification avaiable at most and were identified with the same Family were aggregated to a taxon representative of the Family. 664 Genera (or above) taxa were present in the data after aggregation.

For each pair of body sites, each subject had two samples, one in each body site. In order to avoid across sample dependencies only one of the samples per subject, selected at random, was considered for analysis.

Genera which appeared in less than 2.5\% of the subjects were removed. Some samples contained an irregular low number of reads due to technical faults. Therefore,  at each pairwise comparison of body sites,  the 10\% of samples with the lowest number of reads (in sample) were removed. An alternative way to filter technical faults would have been to set a minimal number of counts required of a valid sample. However, sampling locations exhibit different sequencing depths, and that would require a specific cutoff value for technical faults for each body sampling location.

Table~\ref{table:RDE_HMP} describes the number of discoveries in each pairwise comparison of body sites. In general, samples taken from skin sites and the vagina have reads concentrated at a smaller number of OTUs, compared with samples taken from the oral cavity. This can be seen by the number of taxa considered in the comparisons inside the oral cavity compared with comparisons between skin sites. As observed taxa are more abundant in the oral cavity, more differentially abundant taxa are observed in pairwise comparisons by all methods. W-CSS, a method for marginal inference, has more discoveries compared to ANCOM and DACOMP, across most pairwise comparisons. This is not suprising since W-CSS does not control for compositionality. When comparing ANCOM and DACOMP across the oral cavity, many of the discoveries of DACOMP are shared by ANCOM. Most discoveries of  DACOMP are also shared with ALDEx2-t.

Interestingly, in pairwise comparisons of skin sites, some methods discover differentially abundant taxa between the left and right Antecubital fossa and the left and right Retroauricular crease. ANCOM discovers differentially abundant taxa in both comparisons. When comparing the left and right Retroauricular crease W-TSS (not shown in table) discovers 4 taxa as differentially abundant, none of the discoveries are shared with ANCOM. When comparing the left and right Antecubital fossa, DACOMP-ratio (not shown in table) has a single discovery, which one of the two taxa discovered by ANCOM in this comparison. It is likely that these are false positive findings since there is little agreement between the different methods (DACOMP, ALDEx2 and CSS discover no differentially abundant taxa), and there is no plausible reason to have differentially abundant taxa in these pairwise comparisons. This result is in line with our observation of ANCOM’s empirical decision rule to not be valid under the global null, as discussed in \S~\ref{app:ANCOM_global_null}. 

\begin{table}[htbp]
	\captionsetup{width=1.2\textwidth}
	{	\caption{Pairwise comparison of adjacent body sites in the Human Microbiome Project. For each pair of body sites (columns 1-2), the number of taxa (genera or above) considered for differential abundance between the two sites (coloumn 3), the number of differentially abundant taxa discovered by ANCOM, W-CSS,ALDEx2-t and DACOMP (columns 4-7), the number of discoveries shared by ANCOM and DACOMP (column 8) and the size of reference set (column 9). The BH procedure was applied at level $q=0.1$. For DACOMP $S_{crit}$ was set to 1.3.}
		\label{table:RDE_HMP}}
	\begin{adjustwidth}{-2in}{-2in}  
		\footnotesize
		\centering
		
		\begin{tabular}{llrrrrlrr}
			\hline
			Site 1 & Site 2 & NR.Taxa & ANCOM & W-CSS & ALDEx2-t & DACOMP & Shared & $|B|$\\ 
			\hline
			Saliva & Tongue\_dorsum & 111 &  36 &  83 &  26 &  24 &  18 &  67 \\ 
			Saliva & Hard\_palate & 147 &  30 &  46 &  25 &  29 &  24 &  92 \\ 
			Saliva & Buccal\_mucosa & 145 &  49 &  63 &  31 &  31 &  28 & 106 \\ 
			Saliva & Attached\_Keratinized\_gingiva & 138 &  67 &  93 &  47 &  38 &  38 &  93 \\ 
			Saliva & Palatine\_Tonsils & 146 &  39 &  64 &  28 &  41 &  30 &  86 \\ 
			Saliva & Throat & 156 &  25 &  47 &  20 &  16 &  15 & 126 \\ 
			Saliva & Supragingival\_plaque & 123 &  39 &  95 &  39 &  25 &  22 &  84 \\ 
			Saliva & Subgingival\_plaque & 133 &  44 &  86 &  37 &  32 &  27 &  88 \\ 
			Tongue\_dorsum & Hard\_palate & 106 &  29 &  54 &  20 &  36 &  18 &  65 \\ 
			Tongue\_dorsum & Buccal\_mucosa & 102 &  52 &  64 &  36 &  29 &  26 &  70 \\ 
			Tongue\_dorsum & Attached\_Keratinized\_gingiva &  91 &  54 &  59 &  38 &  29 &  27 &  60 \\ 
			Tongue\_dorsum & Palatine\_Tonsils &  98 &  23 &  40 &  17 &  25 &  15 &  53 \\ 
			Tongue\_dorsum & Throat & 110 &  16 &  54 &   7 &  14 &  10 &  67 \\ 
			Tongue\_dorsum & Supragingival\_plaque & 101 &  60 &  68 &  43 &  34 &  32 &  61 \\ 
			Tongue\_dorsum & Subgingival\_plaque & 102 &  67 &  76 &  50 &  40 &  36 &  55 \\ 
			Hard\_palate & Buccal\_mucosa & 142 &  38 &  53 &  28 &  39 &  32 &  87 \\ 
			Hard\_palate & Attached\_Keratinized\_gingiva & 137 &  51 &  74 &  33 &  46 &  42 &  81 \\ 
			Hard\_palate & Palatine\_Tonsils & 131 &  29 &  52 &  26 &  35 &  18 &  84 \\ 
			Hard\_palate & Throat & 149 &  37 &  36 &  16 &  20 &  19 & 122 \\ 
			Hard\_palate & Supragingival\_plaque & 119 &  59 &  87 &  47 &  34 &  27 &  83 \\ 
			Hard\_palate & Subgingival\_plaque & 126 &  55 &  83 &  45 &  36 &  35 &  80 \\ 
			Buccal\_mucosa & Attached\_Keratinized\_gingiva & 125 &  36 &  60 &  16 &  32 &  31 &  76 \\ 
			Buccal\_mucosa & Palatine\_Tonsils & 129 &  48 &  60 &  34 &  31 &  26 &  91 \\ 
			Buccal\_mucosa & Throat & 146 &  49 &  58 &  34 &  25 &  22 & 114 \\ 
			Buccal\_mucosa & Supragingival\_plaque & 115 &  40 &  73 &  28 &  30 &  24 &  84 \\ 
			Buccal\_mucosa & Subgingival\_plaque & 127 &  42 &  72 &  33 &  32 &  30 &  87 \\ 
			Attached\_Keratinized\_gingiva & Palatine\_Tonsils & 117 &  51 &  56 &  34 &  30 &  27 &  79 \\ 
			Attached\_Keratinized\_gingiva & Throat & 143 &  48 &  56 &  33 &  28 &  27 & 112 \\ 
			Attached\_Keratinized\_gingiva & Supragingival\_plaque & 101 &  47 &  64 &  31 &  32 &  27 &  66 \\ 
			Attached\_Keratinized\_gingiva & Subgingival\_plaque & 116 &  50 &  66 &  37 &  31 &  30 &  81 \\ 
			Palatine\_Tonsils & Throat & 145 &  17 &  19 &   3 &  13 &  10 & 117 \\ 
			Palatine\_Tonsils & Supragingival\_plaque & 106 &  50 &  67 &  40 &  30 &  26 &  65 \\ 
			Palatine\_Tonsils & Subgingival\_plaque & 120 &  50 &  62 &  42 &  38 &  36 &  72 \\ 
			Throat & Supragingival\_plaque & 150 &  57 &  91 &  39 &  36 &  31 & 110 \\ 
			Throat & Subgingival\_plaque & 144 &  69 & 108 &  50 &  42 &  38 &  98 \\ 
			Supragingival\_plaque & Subgingival\_plaque & 103 &  30 &  44 &  19 &  17 &  12 &  64 \\ 
			Right\_Antecubital\_fossa & Left\_Retroauricular\_crease & 244 &   5 &  50 &   2 &  17 &   1 & 244 \\ 
			Right\_Antecubital\_fossa & Right\_Retroauricular\_crease & 190 &   6 &  44 &   2 &   0 &   0 & 188 \\ 
			Right\_Antecubital\_fossa & Left\_Antecubital\_fossa & 286 &   2 &   0 &   0 &   0 &   0 & 269 \\ 
			Right\_Antecubital\_fossa & Anterior\_nares & 209 &  19 &  50 &   7 &  10 &   7 & 190 \\ 
			Left\_Retroauricular\_crease & Right\_Retroauricular\_crease & 172 &   1 &   0 &   0 &   0 &   0 & 166 \\ 
			Left\_Retroauricular\_crease & Left\_Antecubital\_fossa & 198 &   5 &  81 &   2 &   1 &   0 & 196 \\ 
			Left\_Retroauricular\_crease & Anterior\_nares & 202 &   8 &  11 &   7 &  14 &   7 & 183 \\ 
			Right\_Retroauricular\_crease & Left\_Antecubital\_fossa & 200 &   7 &  54 &   2 &   1 &   1 & 198 \\ 
			Right\_Retroauricular\_crease & Anterior\_nares & 200 &   8 &  15 &   9 &  15 &   6 & 180 \\ 
			Left\_Antecubital\_fossa & Anterior\_nares & 209 &  11 &  54 &   8 &   7 &   6 & 189 \\ 
			Vaginal\_introitus & Posterior\_fornix & 120 &   5 &   9 &   2 &   2 &   0 & 105 \\ 
			Vaginal\_introitus & Mid\_vagina & 129 &   2 &   1 &   0 &   0 &   0 & 106 \\ 
			Posterior\_fornix & Mid\_vagina &  96 &   5 &   5 &   0 &   0 &   0 &  89 \\ 
			\hline
		\end{tabular}
	\end{adjustwidth}
\end{table}

\section{Example for testing for differential abundance against a continuous trait }\label{app:RDE_DILUTION}
This section demonstrates how the DACOMP method can be used to test taxa for differential abundance with respect to a continuous trait. The dataset presented in this section is taken from the study of \citet{stammler2016adjusting}. \citet{stammler2016adjusting} proposed a method for normalizing 16S counts data using a spike-in of bacteria (as opposed to a spike-in of synthetic DNA, as in \citet{quinn2018field}). The method consists of cultivating species of bacteria which are not endemic to the sample being sequenced. These bacteria are inserted into the gathered samples in known amounts, prior to PCR amplifications. After the PCR amplification and amplicon sequencing, spiked-in bacteria are measured alongside bacteria endemic to the measured sample. The authors suggest multiplying each OTU counts vector by the ratio between the absolute number of 'spike-in' bacteria for a single taxon and the number of sequences read for the same taxon. \citet{stammler2016adjusting} compare the reconstructed absolute abundances of 'spike-in' taxa not used for normalization to their known absolute abundances, and show that absolute abundances reconstructed by their proposed method enjoy a higher correlation with the absolute abundance of taxa, compared with absolute abundances reconstructed by other methods.


Regardless of the use of 'spiked-in' bacteria as a means for normalization, the study of \citet{stammler2016adjusting} describes an experiment where the the absolute abundance of 'spiked-in' bacteria changes between samples in a known manner, which is independent of the absolute abundance of other bacteria. Hence, the role of the spiked-in bacteria can be reversed: instead of using the known absolute abundance of the 'spike-in's for normalization, we can check which of the bacteria are differentially abundant with respect to the absolute abundance of 'spike-in's (measured independently of 16S samples). Since the total microbial load 'spiked-in' is independent of the original microbial load, only the 'spiked-in' bacteria are differentially abundant, by the experimental design. One can think of 'reversed experimental design' as 'generating a disease' that affects the absolute abundance of only three taxa. The exogenus magnitude of the 'spike-in' sets the 'magnitude' of the disease, with only three of the absolute abundances taxa of taxa (the 'spike-in's) correlated to the disease.

The experimental design was as follows. Thirty six samples were generated by diluting a fixed mass of a fecal sample by one of the following ratios: 1:1, 1:2.15, 1:3.75, 1:6.53, 1:11.37, and 1:19.82 (six samples generated for each dilution ratio). Next, three types of bacteria were 'spiked-in'. Bacteria from the species \textit{Salinibacter ruber} were inserted in identical amounts to all samples. Two other species, \textit{Rhizobium radiobacter} and \textit{Alicyclobacillus acidiphilus} were inserted in quantities such that the product of absolute abundances of both species was fixed to $2.43\cdot 10^{16}$ counts. These bacterias were inserted in six different log-ratios: -5.49, -3.3, -1.1, 1.1, 3.3, 5.49. Overall, for each of the six dilution ratios, and ratios of \textit{Rhizobium radiobacter} and \textit{Alicyclobacillus acidiphilus}, a single sample was generated (36 samples total). For each of the samples, microbial load measurements for each of 'spiked-in' species were taken, via optical measurements \citet{stammler2016adjusting}.

Of the 36 samples, two samples did not have valid microbial load measurements for the spiked-in taxa, and were removed from the study. The remaining samples had 1775 OTUs remaining and 7593-26091 reads, with a median sequencing depth of 17368 reads. We conducted tests of differential abundance between the microbial counts data and three continuous variables: the dilution factor, having six distinct values given by the experimental design; the microbial load measurement of the 'spiked-in' \textit{Rhizobium radiobacter}; and the microbial load measurement of the  \textit{Alicyclobacillus acidiphilus} 'spike-in'. 

The method used for testing for differential abundance are: DACOMP, with a Spearman rank - correlation test, as described in \S~\ref{sec:valid_approach} ; ALDEx2, using the argument \verb|test = 'glm'| that allows testing for differential abundance with respect to a continuous condition; Spearman rank-correlation tests between the a continuous variable and either TSS or CSS transformed counts; and Spearman rank-correlation tests between the continuous variable and CLR transformed counts. The CLR transformation was done either with a pseudocount of either 1 or 0.5 (two test variants) in order to avoid division by zero. Adjustment for multiplicity was done with the BH correction at level $q=0.1$.

\begin{figure}[htbp]
	
	\centering
	\includegraphics[width=1.0\textwidth]{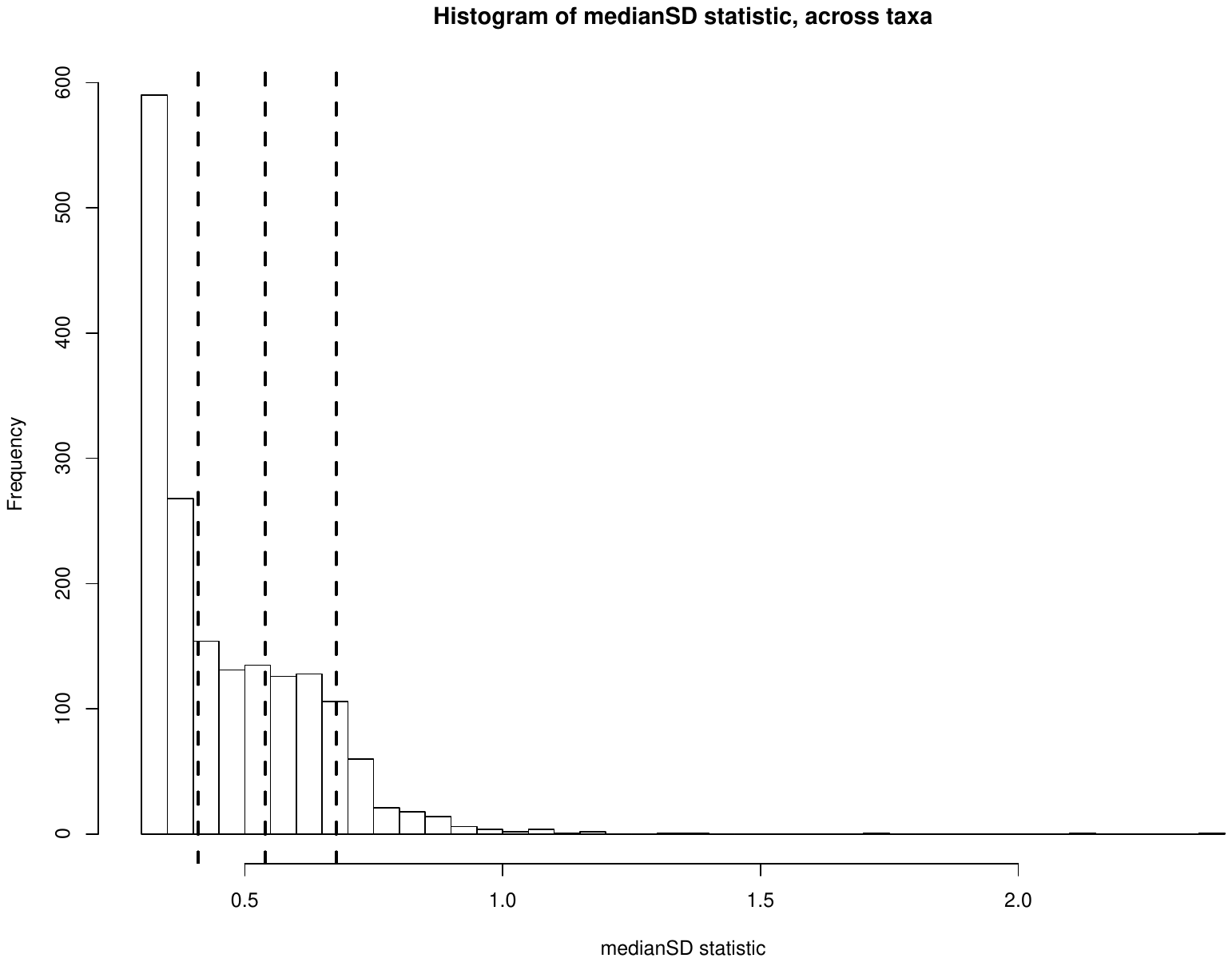}
	\caption{Distribution of the $median\,SD$ statistic for the 1775 taxa in the dilution experiment example described in \S~\ref{app:RDE_DILUTION}.
	}
	\label{figure:app_RDE_Dilution_ref_select}
\end{figure}

References for the DACOMP procedure were selected using the $median\,SD$ statistic, described in \S~\ref{subsec:choosing_reference}. For this example, $S_{crit}$ was set to $0.5$, and we required that the number of reads available under the reference set would be in the range $\left[20,200\right]$. Figure~\ref{figure:app_RDE_Dilution_ref_select} visualizes the empirical distribution of the $median\,SD$ statistic across the different taxa. The value of 0.5 is demonstrated to be between the 2nd and 3rd quartiles, and hence is a valid choice for the experimental setup described, where the percentage of differentially abundant taxa is known to be low.

Table~\ref{app:table_Dilution_Spikein} shows the number of 'spike-in' taxa discovered as differentially abundant. When the variable tested for differential abundance is the microbial load measurement of either \textit{Rhizobium radiobacter} or \textit{Alicyclobacillus acidiphilus}, both taxa are found to be differentially abundant by all methods, as they are negatively correlated by the experimental design. When taxa are tested for differential abundance against the dilution factor,  DACOMP, ALDEx2 and TSS normalization find \textit{Salinibacter ruber} to be associated with the dilution factor, as expected. Tests based on CSS normalization find no taxa to be differentially abundant. Tests based on CLR normalization find two of the spiked-in taxa to be differentially abundant, however, this has to do providing marginal inference alone, rather than increased statistical power.

Table~\ref{app:table_Dilution_NonSpikein} shows the number of non 'spike-in' taxa discovered for all methods. When testing for differential abundance against the dilution factor, CLR and TSS based methods report an exceedingly high number of taxa discovered as differentially abundant, as they provide marginal inference alone, see Example 1 in \S~\ref{sec:Introduction}.  Interestingly, when testing for differential abundance against the microbial load measurements of the two spike-ins, an additional taxon is found to be differentially abundant by all methods but ALDEx2. This taxon is identified as uncultured bacteria from the family \textit{Phyllobacteriaceae} of the order \textit{Phyllobacteriaceae} (the same Order as \textit{Rhizobium radiobacter}). Counts from this taxon are highly correlated to the counts of \textit{Rhizobium radiobacter} ($P$-value = $5.6\cdot 10^{-7}$, Spearman rank-correlation). Considering the nature of the experimental design, this discovery shared by all methods could be a bacteria contaminating the sample of \textit{Rhizobium radiobacter}. Under the premise that this taxon is a contamination in the 'spike-in' of \textit{'Rhizobium radiobacter'}, it is interesting to note the it can be detected by DACOMP, that uses rarefaction, and not by ALDEx2. One possible cause could be the reduction in the number of hypotheses tested by DACOMP, since taxa found in the reference set are not tested for differential abundance. 

To summarize, DACOMP is the only method that performs all three: (1) detects the bacteria spiked-in, when testing for differential abundance against a continuous variable, as demonstrated in Table~\ref{app:table_Dilution_Spikein}; (2) provides control of the false positive rate, as demonstrated in table \ref{app:table_Dilution_NonSpikein} ; (3) is able to detected the contaminant taxon as differentially abundant, presented in table \ref{app:table_Dilution_NonSpikein}.

\begin{table}[htbp]
	\caption{Number of 'spike-in' declared differentially abundant, in the dilution experiment of \S~\ref{app:RDE_DILUTION}. Columns 1-6 correspond to the different methods, while different rows correspond to different continuous variables.}	
	\label{app:table_Dilution_Spikein}
	\centering
	\begin{tabular}{rrrrrrr}
		\hline
		& DACOMP & ALDEx2 & CSS & TSS & CLR\_1 & CLR\_0\_5 \\ 
		\hline
		S1 &   2 &   2 &   2 &   2 &   2 &   2 \\ 
		S3 &   2 &   2 &   2 &   2 &   2 &   2 \\ 
		DilutionFactor &   1 &   1 &   0 &   1 &   2 &   2 \\ 
		\hline
	\end{tabular}
\end{table}

\begin{table}[htbp]
	\caption{Number of non 'spike-in' taxa declared differentially abundant, in the dilution experiment of \S~\ref{app:RDE_DILUTION}. Columns 1-6 correspond to the different methods, while different rows correspond to different continuous variables.}	
	\label{app:table_Dilution_NonSpikein}
	\centering
	\begin{tabular}{rrrrrrr}
		\hline
		& DACOMP & ALDEx2 & CSS & TSS & CLR\_1 & CLR\_0\_5 \\ 
		\hline
		S1 &   1 &   0 &   1 &   1 &   1 &   1 \\ 
		S3 &   0 &   0 &   1 &   1 &   1 &   1 \\ 
		DilutionFactor &   0 &   0 &   0 & 246 & 1114 & 1114 \\ 
		\hline
	\end{tabular}
\end{table}

\end{document}